\documentclass[aip,reprint,twocolumn,showpacs,preprintnumbers,amsmath,aps,superscriptaddress,floatfix]{revtex4-1}
\usepackage{graphicx}
\usepackage{amsmath}
\usepackage{bm}
\usepackage{subfigure}
\usepackage{xcolor}

\newcommand{\rev}[1]{\textcolor{black}{{#1}}}
\newcommand{\revminor}[1]{\textcolor{black}{{#1}}}
\definecolor{darkblue}{rgb}{0,0,0.6}
\usepackage[colorlinks,linkcolor=darkblue,citecolor=darkblue,urlcolor=darkblue]{hyperref}

\begin{document}

\title{On the overlap between configurations in glassy liquids}

\author{Benjamin Guiselin} \email{benjamin.guiselin@umontpellier.fr}

\affiliation{Laboratoire Charles Coulomb (L2C), Universit\'e de Montpellier, CNRS, 34095 Montpellier, France}

\author{Gilles Tarjus} 

\affiliation{LPTMC, CNRS-UMR 7600, Sorbonne Universit\'e, 4 Pl. Jussieu, 75252 Paris cedex 05, France}

\author{Ludovic Berthier} 

\affiliation{Laboratoire Charles Coulomb (L2C), Universit\'e de Montpellier, CNRS, 34095 Montpellier, France}

\affiliation{Department of Chemistry, University of Cambridge, Lensfield Road, Cambridge CB2 1EW, United Kingdom}

\date{\today}

\begin{abstract}
The overlap, or similarity, between liquid configurations is at the core of the mean-field description of the glass transition, and remains a useful concept when studying three-dimensional glass-forming liquids. In liquids, however, the overlap involves a tolerance, typically of a fraction $a/\sigma$ of the inter-particle distance, associated with how precisely similar two configurations must be for belonging to the same physically relevant ``state''. Here, we systematically investigate the dependence of the overlap fluctuations and of the resulting phase diagram when the tolerance is varied over a large range. We show that while the location of the dynamical and thermodynamic glass transitions (if present) is  independent of $a/\sigma$, that of the critical point associated with a transition between a low- and a high-overlap phases in the presence of an applied source nontrivially depends on the value of $a/\sigma$. We rationalize our findings by using liquid-state theory and the hypernetted chain (HNC) approximation for correlation functions. In addition, we confirm the theoretical trends by studying a three-dimensional glass-former by computer simulations. We show in particular that \rev{a range of $a/\sigma$ below what is commonly considered maximizes the temperature of the critical point, pushing it up in a liquid region where viscosity is low and computer investigations are easier due to a significantly faster equilibration.}
\end{abstract}

\maketitle

\section{Introduction}

At the mean-field level glass formation from a liquid is described as a {\it bona fide} thermodynamic transition\cite{kirkpatrick1989scaling,wolynes2012structural,parisi2020theory}. An order parameter can then be identified and, whereas several choices are possible\cite{singh1985hard,wolynes2012structural,parisi2020theory}, one which has proven efficient for systematic investigations is the similarity or overlap between liquid configurations. From the large body of work produced in this direction, it is now understood that the notion of overlap allows one to characterize the statistical properties of the underlying free-energy landscape\cite{monasson1995structural,franz1995recipes,franz1998effective,franz1997phase, cardenas1999constrained,cardenas1998glass,mezard1999compute,cavagna2009supercooled}, the thermodynamics of the ideal glass phase\cite{PhysRevLett.82.747,mezard1999first,cammarota2012ideal}, the dynamical glass transition\cite{kirkpatrick1987connections,kirkpatrick1987dynamics,franz2011field}, the configurational entropy\cite{franz1997phase,berthier2014novel,berthier2019configurational}, etc. Beyond the mean-field description, the spatial fluctuations of the overlap can also be studied and give access to characteristic length scales, such as the point-to-set length\cite{bouchaud2004adam,franz2007analytic, biroli2008thermodynamic,berthier2012static, nagamanasa2015direct, yaida2016point, berthier2017configurational}, and effective field-theoretical models of glassy liquids\cite{franz2011field,dzero2009replica,dzero2005activated,franz2013universality, biroli2014random, rizzo2016dynamical,biroli2018random1,biroli2018random2}. The recognition of the overlap between configurations as a key quantity for glassy systems comes from spin-glass theory\cite{mezard1987spin}. It has been fully developed within the replica formalism where the overlap quantifies the correlation between distinct replicas, correlations that reflect the properties of the free-energy landscape and the existence of multiple metastable states. 

For lattice models, the similarity or overlap between configurations is naturally described by considering an on-site variable: {\it e.g.} for an Ising spin glass, one considers at each lattice site the product of the spins in two configurations; one can further average this product over the whole sample to obtain a global measure of the similarity between the two configurations, taking in this case values between $-1$ for complete anti-correlation to $+1$ for complete correlation\cite{mezard1987spin}. (A slightly different quantity, the bond overlap which considers nearest-neighbor pairs, has also been analyzed\cite{mezard1987spin}.) For liquids, and more generally particle systems in the continuum, the definition requires a little more insight: one should account for (i) permutations of identical particles and (ii) the fact that at a nonzero temperature particles in two similar configurations never sit exactly at the same place, as already illustrated by two distinct thermal configurations of the same ideal crystal. The first point is straightforwardly implemented but the second one requires the introduction of a tolerance that takes two configurations as similar if the particle centers in the two configurations differ by at most a small but nonzero distance to be fixed by some physical arguments\cite{mezard1999first,coluzzi2000thermodynamical,berthier2013overlap,berthier2014novel,berthier2015evidence}. In a dense liquid, as in a solid, it is reasonable to identify this distance with the typical length associated with vibrational motions, a length which is a fraction of the interparticle distance. In concrete terms, considering a single-component atomic liquid for simplicity, the overlap between two configurations $\alpha$ and $\gamma$ of $N$ atoms in a volume $V$, ${\bf r}_\alpha^N\equiv \{{\bf r}_{\alpha,i}\}$ and ${\bf r}_\gamma^N\equiv \{{\bf r}_{\gamma,i}\}$, is defined as
\begin{equation}
\label{eq_definitionQ}
Q_a[\mathbf r_\alpha^N,\mathbf r_\gamma^N]=\frac 1N \sum_{i,j=1}^N w(\vert{\bf r}_{\alpha,i}-{\bf r}_{\gamma,j}\vert/a),
\end{equation}
with $a$ a fraction of the typical interatomic distance $\sigma$ and $w(x)$ a step function or a smooth variant of it which is $1$ for $x<1$ and $0$ for $x>1$\footnote{Note that in a related procedure, the overlap can be defined by first discretizing space\cite{cammarota2010phase}: the sample is divided in small boxes with a linear size $a$ of the order of a fraction of the inter-particle distance and a discrete variable is introduced in each box that takes the value $1$ if a particle center is present and $0$ otherwise; the overlap then uses the product of these pseudo-on-site variables in two different configurations and the tolerance is now associated with the box size $a$.}. The double sum in Eq.~(\ref{eq_definitionQ}) takes care of particle permutations. In all previous studies on model glass-forming liquids, the cutoff $a$ has been taken such that $a/\sigma=0.2-0.3$, which seems a physically plausible value for a typical vibrational length. However, no one has so far investigated what the effect of changing the ratio $a/\sigma$ over a significant range is. The goal of the present work is to fill this gap.

\rev{The overlap we consider is a static quantity, with no reference to the dynamics. One can also investigate the similarity between a configuration at a given time $t^\prime$, and the same configuration, having evolved under the dynamics of its constituents, after an elapsed time $t$\cite{parisi1997short}. In Eq.~(\ref{eq_definitionQ}), ${\bf r}_{\alpha,i}$ and ${\bf r}_{\gamma,j}$ are then replaced by ${\bf r}_{i}(t^\prime)$ and ${\bf r}_{j}(t+t^\prime)$. The fluctuations of this time-dependent overlap, as quantified by a generalized dynamical susceptibility often referred to $\chi_4(t)$ and a $4$-point space-time correlation function, are useful to describe the spatially heterogeneous nature of the dynamics and the growing extent of the dynamical correlations as one cools a glass-forming liquid. In this case, too, the definition of the overlap involves a tolerance $a$, but the physical significance and the effect of the latter are more readily understandable\cite{lavcevic2003spatially}. If $a/\sigma$ is too small, the involved dynamics is controlled by only weakly coupled vibrations and the dynamical correlations remain small, while if $a/\sigma$ is large one encounters the rather unphysical feature that a particle from the configuration at time $0$ can overlap with several other particles at time $t$. In between there is an optimal value of the ratio (around $0.3$) for which the spatial correlations in the dynamics grow bigger. We will no further discuss the dynamic overlap and only study the \textit{static overlap} between configurations sampled from equilibrium distributions.}

We focus on the setting put forward by Franz and Parisi\cite{franz1995recipes} in which one considers the effective potential associated with the typical free-energy cost to constrain an equilibrium liquid configuration ${\bf r}^N$ at a fixed overlap value $Q$ with a reference liquid configuration ${\bf r}_0^N$. To investigate different regions of the free-energy landscape the reference configuration can be drawn from the equilibrium Boltzmann distribution at various temperatures $T_0$ and densities $\rho_0$. \revminor{In most of what follows, we focus on the situation where the reference configuration is taken from the equilibrium distribution at the same temperature $T$ and density $\rho$ as the constrained equilibrium configuration, and we therefore present the formalism for this case (but generalization is straightforward)}. For a single-component liquid with Hamitonian $H[\mathbf r^N]=(1/2) {\sum^\prime}_{i,j=1}^Nv(\vert\mathbf r_i-\mathbf r_j\vert)$, where $v(r)$ is the pair interaction \revminor{(and where the prime denotes that the sum runs over all pairs of particles with $i\neq j$)}, the Franz-Parisi (FP) potential is then defined as
\begin{equation}
\begin{aligned}
\label{eq_franz-parisi}
&-\beta N V_a(Q)= \\& \int \mathrm{d}{\bf r}_0^N \frac{e^{-\beta H[{\bf r}_0^N]}}{Z_0(T)}\, \ln \int \mathrm{d}{\bf r}^N \frac{e^{-\beta H[{\bf r}^N]}}{Z(T\vert {\bf r}_0^N)}\delta(Q_a[\mathbf r^N,\mathbf r_0^N]-Q),
\end{aligned}
\end{equation}
where $\beta=1/(k_BT)$ with $k_B$ the Boltzmann constant, $Q_a[\mathbf r^N,\mathbf r_0^N]$ is defined in Eq.~(\ref{eq_definitionQ}), and $Z$ and $Z_0$ are normalization factors ({\it i.e.} partition functions). We have added a subscript $a$ on the potential to recall that its definition depends on the parameter $a$.

In mean-field treatments of glass formation, as well as in mean-field (exact) models and in liquids in infinite dimension, the FP potential plays the role of a Landau free-energy function of the order parameter\cite{franz1995recipes,franz1997phase,franz1998effective,cardenas1999constrained,cardenas1998glass}. \rev{It contains essential information on the statistical properties of the free-energy landscape of the glass-former.} The FP potential always has a minimum corresponding to decoupled replicas with a small overlap. For a low enough temperature or a high enough density, \rev{the relevant region of phase space splits into an exponentially large number of metastable states, and }a second, metastable, minimum corresponding to coupled replicas and a higher overlap appears. \rev{The difference in potential between the metastable and the stable minima represents the free-energy cost to constrain the system within a metastable state selected by the reference configuration, and hence it provides a direct measure of the ``configurational entropy'' (or ``complexity'') of the system, which represents the logarithm of the number of metastable states divided by $N$.} When $T$ is further decreased, the second minimum deepens and reaches the same free energy as the decoupled minimum at a temperature $T_K$ where the thermodynamic glass transition (random first-order transition\cite{kirkpatrick1989scaling}\revminor{, often referred to as the ``Kauzmann transition'' in the literature}) takes place. As will be discussed in more detail below, the two critical temperatures (or densities) $T_d$, at which the metastable minimum first appears and which corresponds to the ``dynamical glass transition'' and to the ``spinodal'' of the high-overlap phase, and $T_K$ are independent of the choice of $a$. On the other hand, at higher temperature than $T_d$ the potential $V_a(Q)$ retains some nonconvex features that only disappear at a temperature $T_c$, which, as we will show in a mean-field approximate liquid theory (hypernetted-chain or HNC\cite{hansen1990theory}) depends on the cutoff parameter $a$ in a nontrivial way. 

\begin{figure}[t]
\centering
\includegraphics[width=.494\columnwidth]{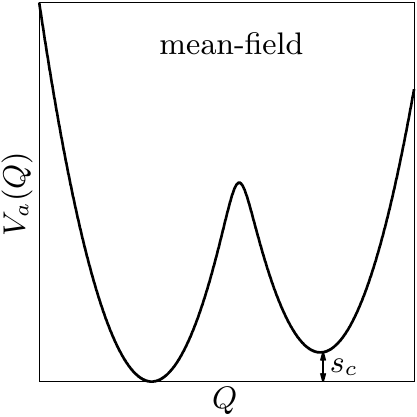}
\includegraphics[width=.494\columnwidth]{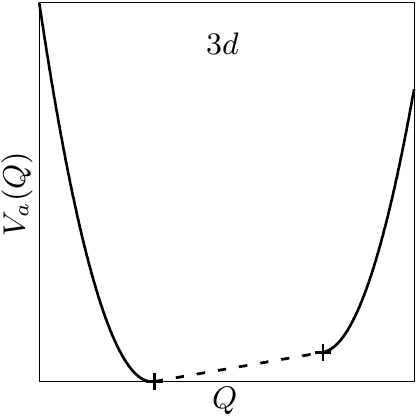}
\includegraphics[width=.494\columnwidth]{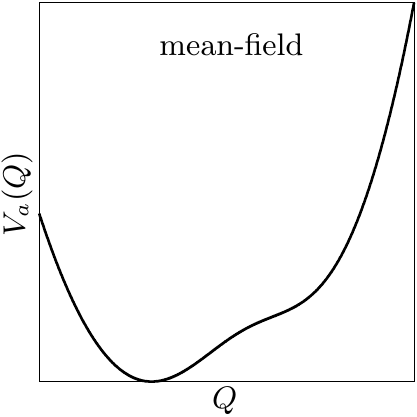}
\includegraphics[width=.494\columnwidth]{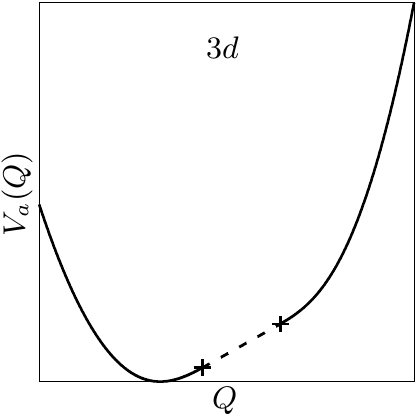}
\caption{\rev{Sketch of the Franz-Parisi potential $V_a(Q)$ in the mean-field description (left panels) and for a three-dimensional glass-former (right panels). In the top panels the temperature is slightly above $T_K$, at which a thermodynamic phase transition takes place: a second minimum is present in the mean-field description and $s_c$ is the configurational entropy; the finite-dimensional system has a convex potential but with a linear segment between a low-overlap point and a high-overlap one. In the bottom panels the temperature is slightly below $T_c$, at which a singular point with $V_a''(Q)=V_a'''(Q)=0$ exists: no second minimum remains in mean-field but the potential is still nonconvex; in finite dimensions the size of the linear segment has shrunk with increasing temperature and goes to zero at $T_c$. Above $T_c$ the FP potential is convex with $V_a''(Q)>0$ everywhere, in both mean-field and finite dimensions.}}
\label{Fig_sketch}
\end{figure}

\rev{When transferred to $3$-dimensional glass-forming liquids, the mean-field scenario can no longer hold as such. Spatial fluctuations of all local quantities, including the local overlap, brought about by the finite-dimensional nature of space, radically change notions such as metastability and spinodal. A ``metastable'' state different from the stable one can then only be defined over a restricted time\footnote{\revminor{The supercooled liquid state is of course metastable with respect to the crystal but its lifetime is much larger than the $\alpha$-relaxation time or local equilibration time in the liquid. This is a well-documented phenomenon and it is very different from metastable glassy states for which both lifetime and local equilibration time are controlled by the same $\alpha$-relaxation time.}}, and the limit of stability of such a state, its spinodal, is then blurred and can at best remain as a crossover. For this reason, no second minimum in the FP potential $V_a(Q)$ and no dynamical transition at $T_d$ can be found in $3d$ liquids in the thermodynamic limit. Similarly, the FP potential should remain convex at all temperatures. This however does not prevent the appearance of singularities in this potential, in the form of straight segments\cite{biroli2016role}: see Fig.~\ref{Fig_sketch} for a sketch. A first-order-like glass transition at a Kauzmann temperature $T_K$ is still possible, as is possible the existence at higher temperature of a straight-line portion that shrinks as one increases the temperature and disappears at a temperature $T_c$ at which $V_a(Q)$ has vanishing second and third derivatives.}

\rev{To unfold these singular features of the FP potential, which, we recall, characterize the properties of the liquid landscape in configurational space, it is convenient to apply a source $\epsilon$ linearly coupled to the overlap $Q_a$. Singularities of $V_a(Q)$, in the form of either a nonconvex portion in the mean-field description  or of a straight segment in finite-dimensional systems in the thermodynamic limit (see Fig. \ref{Fig_sketch}) lead to a line of first-order transition between a low-overlap phase and a high-overlap phase in the $\left(\epsilon,~T\right)$ diagram. This line emerges from the Kauzmann transition point $T_K$ at zero coupling $\epsilon$\cite{franz1995recipes,franz1998effective,franz1997phase,cardenas1999constrained,cardenas1998glass} and terminates in a critical point located exactly at the temperature $T_c$, but at a nonzero critical coupling $\epsilon_c$. We stress again that, contrary to the dynamical transition at $T_d$, this whole line from $T_K$ to $T_c$ may {\it a priori} be meaningful beyond mean-field. The existence or not of such a nontrivial extended phase diagram in actual $3$-dimensional glass-forming liquids is then a key test for the practical relevance of the mean-field scenario of the glass transition. What we stress in this work is that the position of the first-order transition line and of the critical endpoint in the phase diagram depends on the choice of the cutoff parameter $a$. We study its variation in detail and discuss the consequences.}


The rest of the paper is organized as follows. In Sec.~\ref{sec:statmech}, we present the general statistical-mechanical framework to describe a transition from low-overlap to high-overlap phases starting from liquid-state theory. In particular, we clarify the dependence of several quantities on the cutoff parameter $a$ entering in the definition of the overlap. In Sec.~\ref{sec:HNC}, we present the HNC approximation as a mean-field-like closure of the theory developed in the previous section. Within this approximation, we obtain equations that can be solved numerically to study quantitatively the influence of the parameter $a$. The results concerning more specifically the critical endpoint of the line of first-order transition between phases of low and high overlap are presented in Sec.~\ref{sec:HNC_results}. We also present analytical arguments and a detailed discussion for the behavior in the limiting cases of small and large values of $a$. In Sec.~\ref{sec:simulation}, we give results of a computer simulation of a three-dimensional model glass-forming liquid and we show that they  corroborate our theoretical analysis. Finally, we \revminor{discuss the implications of our study and} conclude in Sec.~\ref{sec:conclusions}. Additional details are given in Appendix~\ref{app_analysisHNC}.

\section{Statistical mechanics of glass-forming liquids}

\label{sec:statmech}

The most convenient way to compute the FP potential defined in Eq.~(\ref{eq_franz-parisi}) is to introduce $n$ replicas of the constrained equilibrium configuration $\mathbf r_1^N, \cdots, \mathbf r_n^N $, in order to replace the logarithm appearing in the definition by a more tractable expression and to take at the end the limit $n\to 0$. As is also standard, in the spirit of the equivalence between the canonical and the grand-canonical equilibrium ensembles in the thermodynamic limit, one can replace the ensemble in which $Q$ is the control parameter by an ensemble in which it is the conjugate source $\epsilon$ that is the control parameter. This replacement amounts to a Legendre transform\cite{franz1995recipes,franz1997phase,franz1998effective,cardenas1999constrained,cardenas1998glass},
\begin{equation}
\label{eq_Legendre}
N \beta V_a(Q)= N \beta F_a(\epsilon) + N \beta \epsilon \, Q,
\end{equation}
with
\begin{equation}
\label{eq_epsilon-Q}
\beta\epsilon=\beta V'_a(Q),
\end{equation}
where a prime denotes a derivative with respect to the argument.

Within this replica formalism one is led to consider an equilibrium liquid mixture of $n+1$ components with Hamiltonian
\begin{equation}
\begin{aligned}
 \label{eq_replica_hamiltonian}
& H_{\rm rep}[\{\mathbf r_\alpha^N\}]= \frac 12 \sum_{\alpha,\gamma=0}^n\sum_{i,j=1}^N w_{\alpha\gamma}(\mathbf{r}_{\alpha,i},\mathbf{r}_{\gamma,j}\vert \epsilon,a)  ,
\end{aligned}
\end{equation}
where the interaction potentials are given by
\begin{equation}
\begin{aligned}
 \label{eq_w}
w_{\alpha\gamma}(\mathbf r,\mathbf{r'}\vert\epsilon,a)=&\delta_{\alpha\gamma}\, v(\vert \mathbf r-\mathbf{r'}\vert) 
-[(1-\delta_{\alpha 0})\delta_{\gamma 0}+\\&
\delta_{\alpha 0}(1-\delta_{\gamma 0})]\epsilon \,w(\vert \mathbf r-\mathbf{r'}\vert/a)\,.
\end{aligned}
\end{equation}

A key quantity in liquid-state theory to access the FP potential is the so-called Morita-Hiroike functional $ \Gamma_{MH}$ of the 1- and 2-particle densities\cite{morita1960new,morita1961new} for the replicated $(n+1)$-component liquid mixture, which is obtained via a Legendre transform between the interaction potentials $w_{\alpha\gamma}$ and the 2-particle densities $\rho_{\alpha\gamma}^{(2)}$. Since we are interested in homogeneous phases, it is sufficient to consider translationally invariant densities; moreover, all replicas have the same 1-particle density $\rho$. After introducing the total correlation functions $h_{\alpha\gamma}$ via $\rho_{\alpha\gamma}^{(2)}(\mathbf r,\mathbf r')=\rho^2[1+h_{\alpha\gamma}(\vert \mathbf r-\mathbf r'\vert)]$ (where $g_{\alpha\gamma}=1+h_{\alpha\gamma}$ is the conventional pair correlation function)\cite{hansen1990theory}, the Morita-Hiroike functional (per unit volume) reads\cite{biroli2018random1,biroli2018random2,morita1960new,franz2013static}
\begin{equation}
\begin{aligned}
 \label{eq_morita1960new,morita1961new}
&\Gamma_{MH}[\{h_{\alpha\gamma}\};\rho]= \\&(n+1) \rho(\ln \rho-1)+ \frac 1{2}\rho^2\sum_{\alpha\gamma} \int_{\mathbf r}[1+h_{\alpha\gamma}(r)]\beta w_{\alpha\gamma}(r\vert \epsilon,a)\\&
+ \frac 1{2}\rho^2\sum_{\alpha\gamma} \int_{\mathbf r}[1+h_{\alpha\gamma}(r)]\left \lbrace\ln[1+h_{\alpha\gamma}(r)]-1\right\rbrace  \\&
+ \frac 12\sum_{p\geq 3} \frac{(-1)^p\rho^p}{p}\sum_{\alpha_1\cdots \alpha_p}\int_{\mathbf r_2}\int_{\mathbf r_3}\cdots\int_{\mathbf r_p }h_{\alpha_1\alpha_2}(r_2)\times \\& h_{\alpha_2\alpha_3}(\vert \mathbf r_3-\mathbf r_2\vert)\cdots h_{\alpha_p\alpha_1}(r_p) \,+ {\rm 2PI},
\end{aligned}
\end{equation}
with $r=\vert \mathbf r\vert$, $w_{\alpha\gamma}(\vert \mathbf{r}-\mathbf{r'}\vert\,\vert \epsilon,a)=w_{\alpha\gamma}(\mathbf{r},\mathbf{r'}\vert\epsilon,a)$ and where ${\rm 2PI}$ denotes the sum of all 2-particle irreducible diagrams formed with density vertices linked by total correlation functions\cite{morita1960new,morita1961new}. Without these terms the above expression reduces to the well-known HNC approximation of liquid state theory\cite{hansen1990theory}. Note also that the interaction potential with a dependence on $\epsilon$ and $a$ only appears in the second term of the right-hand side of Eq.~(\ref{eq_morita1960new,morita1961new}), so that one can formally rewrite the functional as
\begin{equation}
\begin{aligned}
 \label{eq_intrinsic}
&\Gamma_{MH}[\{h_{\alpha\gamma}\};\rho]= \frac 1{2}\rho^2\sum_{\alpha\gamma} \int_{\mathbf r}[1+h_{\alpha\gamma}(r)]\beta w_{\alpha\gamma}(r\vert \epsilon,a)\\&
+ \mathcal F[\{h_{\alpha\gamma}\};\rho] ,
\end{aligned}
\end{equation}
where $\mathcal F$ is independent of the pair potentials, emphasizing the Legendre transform between the interaction potentials and the 2-particle densities.

The equilibrium total correlation functions are obtained by minimizing the Morita-Hiroike functional,
\begin{equation}
\label{eq_minimization}
\frac{\delta  \Gamma_{MH}}{\delta h_{\alpha\gamma}(r)}=0, \;{\rm or}\; \frac{\delta  \mathcal F}{\delta h_{\alpha\gamma}(r)}=-\frac 12 \rho^2\beta w_{\alpha\gamma}(r\vert \epsilon,a).
\end{equation}
Being interested in the liquid phase above the ideal glass transition and by homogeneous configurations, we can assume replica symmetry between the $n$ constrained replicas (replica $0$ is different due to the attractive coupling) in the solution of the above minimization equations and then take the limit $n\to 0$. One thus needs to consider 4 distinct functions, $h_{11}^*(r)$, $h_{12}^*(r)$, $h_{00}^*(r)$ and $h_{01}^*(r)$, where the superscript $^*$ means that the functions correspond to solutions of the minimization equations\footnote{In mean-field approximations, the minimization equations may of course have several solutions with higher free energy that the global minimum and are then associated to metastable states and saddle-points.}.

We want to focus on the correlation between the constrained replicas and the reference one, {\it i.e.} on $h_{01}(r)$. To do this, one can solve the minimization equations for $h_{00}(r)$, $h_{11}(r)$ and $h_{12}(r)$. The solutions are then functionals of $h_{01}(r)$ and of the potential $v(r)$ [except $h_{00}(r)$ which only depends on $v(r)$ and is decoupled from the other total correlation functions in the limit $n\to 0$]; they depend on $\rho$ but they do not depend on $\epsilon$ and $a$. Let us call $\mathcal F[h_{01};\rho]$ the functional resulting from replacing $h_{00}(r)$, $h_{11}(r)$ and $h_{12}(r)$ in $\mathcal F[\{h_{\alpha\gamma}\};\rho]$ by their solution. Its expression is
\begin{equation}
\begin{aligned}
\mathcal F[h_{01};\rho]&=\lim_{n\rightarrow 0}\left.\left\lbrace \frac{\mathcal F[\{h_{\alpha\gamma}\};\rho]-\mathcal F[h_{00};\rho]}{n}\right\rbrace\right\vert_{\mathrm{RS}}\\&
+\frac{\rho^2}{2}\int_{\mathbf{r}}[1+h_{11}^*(r)]\beta v(r),
\end{aligned}
\end{equation}
with $\mathrm{RS}$ denoting replica symmetry and $\mathcal F[h_{00};\rho]$ the functional for the reference replica only. The key point is that the functional $\mathcal F[h_{01};\rho]$ is independent of $\epsilon$ and $a$. On the other hand, the function $h_{01}^*(r)$ which is now obtained as the solution of
\begin{equation}
\label{eq_minim_h01}
\frac{\delta  \mathcal F[h_{01};\rho]}{\delta h_{01}(r)}=\rho^2\beta \epsilon\, w(r/a)
\end{equation}
depends on $\epsilon$ and $a$. [There is no factor $1/2$ in the expression because we take $h_{10}=h_{01}$ and the change of sign is due to the minus sign in Eq.~(\ref{eq_w}).] However, when $\epsilon=0$, the dependence on $a$ drops out because the right-hand side of the above equation is simply zero.

At this point we can go back to the FP potential $V_a(Q)$. From Eq.~(\ref{eq_definitionQ}) and the definition of $h_{01}(r)$, the overlap $Q_a$ between constrained and reference configurations can be expressed as
\begin{equation}
\label{eq_expressionQ}
Q_a= \rho \int_{\mathbf r}[1+ h_{01}(r)]  w(r/a)  \,.
\end{equation}
When the constrained and the reference configurations are uncorrelated, $h_{01}(r)\equiv 0$, and the overlap takes its  ``random'' value, $Q_{a,{\rm rand}}=\rho \int_{\mathbf r}  w(r/a)$. It is then more convenient to characterize the nontrivial features associated with correlations between replicas through the order parameter
\begin{equation}
\label{eq_expressionDeltaQ}
\Delta Q= Q_a-Q_{a,{\rm rand}}=\rho \int_{\mathbf r} h_{01}(r)  w(r/a) \,.
\end{equation}
The free energy $F_a(\epsilon)$ introduced in Eq.~(\ref{eq_Legendre}) can be derived from the functional $\mathcal F[h_{01};\rho]$ as
\begin{equation}
\label{eq_expression_Free-energy}
\beta F_a(\epsilon)=\frac 1{\rho}\mathcal F[h_{01}^*;\rho]- \rho \int_{\mathbf r}[1+ h_{01}^*(r)] \beta \epsilon w(r/a) , 
\end{equation}
and the FP potential is obtained by the Legendre transform. Expressing it in terms of $\Delta Q$ rather than $Q$, it takes the form 
\begin{equation}
\label{eq_expression_FPpotential}
\beta V_a(\Delta Q)=\frac 1{\rho}\mathcal F[h_{01}^*;\rho]- \rho \int_{\mathbf r} h_{01}^*(r) \beta \epsilon w(r/a) +\beta\epsilon \Delta Q \,,
\end{equation}
where $h_{01}^*(r)$ and $\epsilon$ can now be considered as functions of $\Delta Q$ and $a$.

We are now in a position to discuss two generic properties of the FP potential as a function of the cutoff parameter $a$. 

(1) If the potential has several extrema, as it does in mean-field treatments, the value of $\epsilon$ at these extrema is zero; as stressed above, the function $h_{01}^*(r)$ is then independent of $a$ and corresponds to extrema of the functional $\mathcal F[h_{01};\rho]$. The temperature and density at which these extrema appear and disappear as well as the value of the associated free energy are intrinsic properties of $\mathcal F[h_{01};\rho]$ and therefore do not depend on $a$. {\it As a result, neither $T_d$ nor $T_K$ depend on the choice of $a$.} The value of the overlap at the extrema on the other hand depends on $a$ through Eq.~(\ref{eq_expressionDeltaQ}). Requiring for physical consistency that the value of $\Delta Q$ at the correlated minimum corresponding to the emerging glass phase is positive may put an upper bound on the value of $a$, but this does not correspond to a real physical singularity: this point will be illustrated and discussed in more detail below. In addition, the complexity, which we remind that it represents the free-energy cost to constrain the liquid within a single metastable state and that it corresponds to the height of the secondary minimum in the potential $V_a(Q)$, must also be independent of $a$.

(2) The critical point $T_c$ mentioned in the introduction corresponds to the temperature at which the FP potential either recovers full convexity in mean-field approximations or loses signatures of singular behavior corresponding to the presence of a straight segment in large enough finite-dimensional systems (in finite dimensions the potential is indeed always convex but may display a straight segment between two values of the overlap, see Fig.~\ref{Fig_sketch}). Then, there is a critical value $\Delta Q_c$ at which 
\begin{equation}
\label{eq_critical1}
V''_a(\Delta Q_c)=V'''_a(\Delta Q_c)=0  ,
\end{equation}
and a critical value $\epsilon_c$ such that 
\begin{equation}
\label{eq_critical2}
\epsilon_c=V'_a(\Delta Q_c). 
\end{equation}
From Eq.~(\ref{eq_expression_FPpotential}) one can see that, generically, not only $\Delta Q_c$, but also $\epsilon_c$ and $T_c$ should now depend on $a$. {\it The location of the critical point, and as a consequence of the whole first-order transition line in the $\left(\epsilon,~T\right)$ phase diagram, therefore vary with the choice of $a$.} 

\rev{Let us make a theoretical comment at this point. In the theory of critical phenomena, one is used to distinguish short-range fluctuations due to the microscopic details of a system and long-range, potentially scale-free, fluctuations that appear at criticality. As is well-known in statistical physics and field theory, different  microscopic models may belong to the same universality class at the critical point, hence showing the same long-distance physics. However, the nonuniversal quantities, such as the location of the critical point, depend on the short-range fluctuations as well and vary from one model to another. The situation is more subtle here. From the very same liquid, one may build a family of effective theories for the overlap that is indexed by the tolerance $a$. One could anticipate that the long-distance physics (the universality class of the critical point) is independent of $a$ but that nonuniversal quantities depend on $a$. But this is not the whole story: for instance, the extrema of the FP potential are independent of $a$, as a consequence of the property that the intrinsic generating functional is independent of $a$ and $\epsilon$; and this applies whether or not the system is at criticality.}

In the next section we will illustrate the above described generic features in the case of an approximate mean-field treatment based on the HNC closure.

\section{HNC approximation and the Franz-Parisi potential}
\label{sec:HNC}

\revminor{The HNC approximation is one of the standard tools of liquid-state theory to describe the structure and the thermodynamics of liquids.} It amounts to neglecting all 2-PI diagrams in the Morita-Hiroike functional given in Eq.~(\ref{eq_morita1960new,morita1961new}). The minimization equations in Eqs.~(\ref{eq_minimization}) can be cast in a more familiar form by introducing the direct correlation functions $c_{\alpha\gamma}(r)$ that are related to the total correlation functions by the Ornstein-Zernicke equations\cite{hansen1990theory}. Assuming again replica symmetry in the limit $n\to 0$, one finds in Fourier space
\begin{equation}
\begin{aligned}
 \label{eq_HNC_Fourier}
&1+\rho h_{00}(q)=\frac1{1-\rho c_{00}(q)}\\&
1+\rho h_{{\rm con}}(q)=\frac1{1-\rho c_{{\rm con}}(q)}\\&
h_{12}(q)=[1+\rho h_{{\rm con}}(q)]^2\left\lbrace c_{12}(q)+\rho [1+\rho h_{00}(q)] c_{01}(q)^2\right\rbrace\\&
h_{01}(q)=[1+\rho h_{00}(q)][1+\rho h_{{\rm con}}(q)]c_{01}(q),
\end{aligned}
\end{equation}
where we have introduced the ``connected'' correlation functions, $h_{{\rm con}}=h_{11}-h_{12}$ and $c_{{\rm con}}=c_{11}-c_{12}$\footnote{In the context of disordered systems, $h_{12}$ and $c_{12}$ are also called the ``disconnected'' correlation functions.}, and kept the same notation for the functions in Fourier and in real spaces. The HNC closure derived from the minimization equations can then be written as
\begin{equation}
\begin{aligned}
 \label{eq_HNC_closure}
&c_{00}(r)= -\beta v(r) +h_{00}(r) - \ln[1+h_{00}(r)]\\&
c_{11}(r)=-\beta v(r) +h_{11}(r)- \ln[1+h_{11}(r)]\\&
c_{12}(r)=h_{12}(r) - \ln[1+h_{12}(r)]\\&
c_{01}(r)=\beta\epsilon w(r/a) + h_{01}(r) - \ln[1+h_{01}(r)].
\end{aligned}
\end{equation}
From the solution of these equations, one obtains the free energy $\beta F_a(\epsilon)$ [see Eq.~(\ref{eq_expression_Free-energy})] and then the FP potential [see Eq.~(\ref{eq_expression_FPpotential})], whereas the overlap difference with the random limit $\Delta Q$ is given by Eq.~(\ref{eq_expressionDeltaQ}).

The HNC approximation is of mean-field character as it leads to a nonconvex potential at low enough temperature for glass-forming liquids and then sustains infinitely long-lived metastable states. It has already been well studied in the context of the glass transition\cite{mezard1996tentative,cardenas1999constrained,cardenas1998glass,bomont2014probing,bomont2017coexistence,bomont2015hypernetted,parisi2010mean}, including a calculation of the FP potential\cite{cardenas1999constrained,cardenas1998glass}. Our purpose here is not to repeat all of these calculations but to investigate the role of the cutoff parameter $a$ used in the definition of the overlap.

\begin{figure}[t]
\centering
\includegraphics[width=\columnwidth]{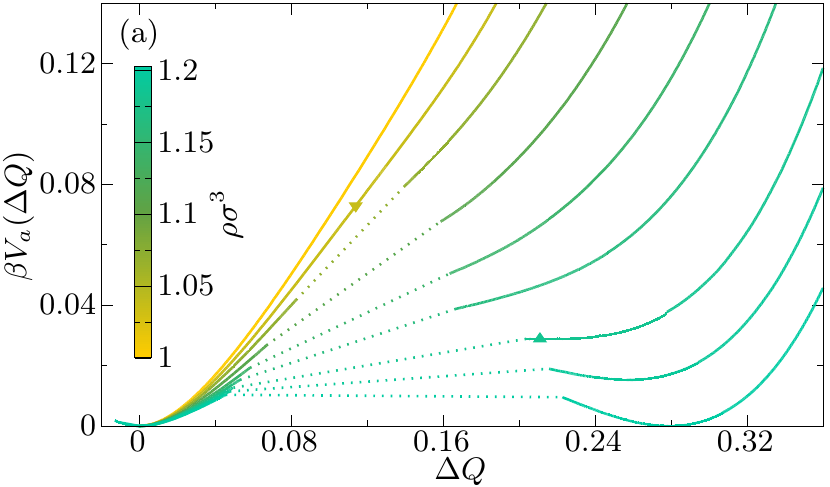} 
\includegraphics[width=\columnwidth]{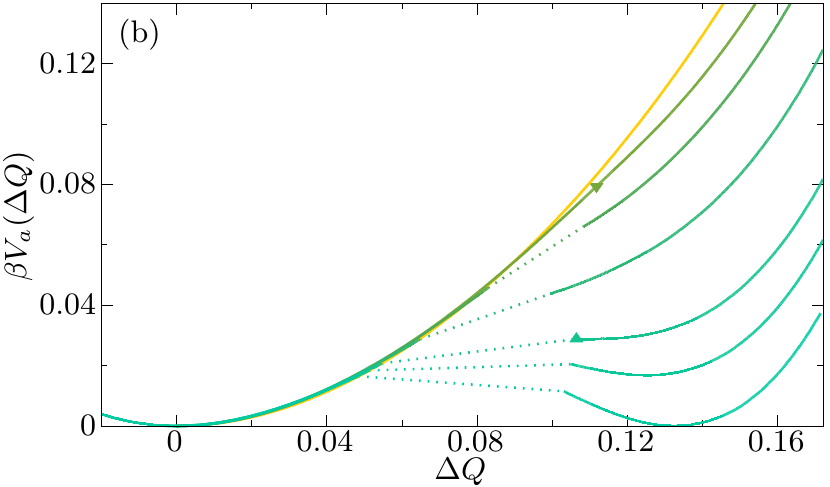} 
\caption{Evolution with density of the Franz-Parisi potential $V_a(\Delta Q)$ in the HNC approximation for a three-dimensional hard-sphere system and two different values of the cutoff parameter $a$: (a) $a/\sigma=0.2$; (b) $a/\sigma=0.5$. In all panels, the up and down triangles mark the values of $\Delta Q$ at the dynamical transition (spinodal of the metastable glass minimum) and the critical point, respectively. \revminor{For both figures, the color code is the same and given by the colorbar in panel (a).} The dotted lines represent the region where there is no replica-symmetric solution to the HNC equations\rev{\cite{Note5}}.}
\label{Fig_FPpotential}
\end{figure}

We consider two different single-component liquid models in three dimensions: a hard sphere model, with $v(r)=0$ for $r\geq\sigma$ and $=\infty$ otherwise, and a soft sphere model, with $v(r) = v_0 [(\sigma/r)^{12} + \kappa_0 + \kappa_2(r/\sigma)^2 + \kappa_4 (r/\sigma)^4]$ for $r<1.25\sigma$ and $v(r)=0$ otherwise, where $v_0$ is the energy scale (the Boltzmann constant $k_B$ is set to unity) and $\kappa_{2l}$ ($l=0,1,2$) are constants that ensure that the potential $v(r)$ and its first two derivatives are continuous in $r=1.25\sigma$. The control parameter is density in the former case and temperature in the latter \revminor{(in this case, the density is set to unity)}. For the threshold function involved in the definition of the overlap [see Eqs.~(\ref{eq_definitionQ}) or (\ref{eq_expressionQ})], we have chosen a continuous one, $w(x)={\rm exp}(-x^4 \ln 2)$. Note that in the HNC approximation where we consider homogeneous configurations, one does not have to worry about crystallization and the liquid always forms an ideal glass through a thermodynamic phase transition at a low-enough temperature $T_K$ or a high-enough density $\rho_K$.

Eqs.~(\ref{eq_HNC_Fourier}) and (\ref{eq_HNC_closure}) are solved iteratively by using a real-space linear mesh of size $\mathrm{d}r=\sigma/128$ for $a\geq 0.1$ and $\mathrm{d}r=\sigma/512$ otherwise (to ensure that $\mathrm{d}r\times a>10$), with a large-distance cutoff of $L=8\sigma$. We have checked that taking a larger cutoff distance and/or a smaller mesh size only leads to very small quantitative change of our results. For a given value of $a$ and a given density $\rho$ (in the hard-sphere case) or a given temperature $T$ (in the soft-sphere case), we compute the curves $\Delta Q^{(\pm)}(\epsilon)$ from Eq.~(\ref{eq_expressionDeltaQ}) by increasing the source $\epsilon$ from $0$ ($\Delta Q^{(+)}$) or decreasing it from a high-enough value ($\Delta Q^{(-)}$). The first-order transition region is detected when there is a range of $\epsilon$ values for which $\Delta Q^{(+)}(\epsilon)\neq \Delta Q^{(-)}(\epsilon)$\footnote{\revminor{The resolution of the HNC equations within the replica-symmetric (RS) formalism is done first fixing the value of $\epsilon$ and then iterating Eqs.~(\ref{eq_HNC_Fourier}) and (\ref{eq_HNC_closure}) until the correlation functions converge. Eventually, the corresponding values of the overlap $Q_a$ and of the Franz-Parisi potential $V_a(Q)$ are calculated. We stress that we do not impose the value of the overlap for the resolution of the equations, but $\epsilon$ instead. As a result, for any value of $\epsilon$, there is always at least one solution of the RS HNC equations. However, RS solutions only correspond to a limited range of overlap values, thus explaining why there is a straight line in the plot of the Franz-Parisi potential (see Figs.~\ref{Fig_FPpotential} and \ref{Fig_diagram-large-a}). We expect that breaking the symmetry between the replicas (a procedure known as RSB) and solving the HNC equations within the RSB formalism would enable us, for intermediate values of $\epsilon$, to sample values of the overlap which are forbidden within the RS formalism.}}. With this procedure, we are able to locate the critical point with an arbitrary degree of precision. In the following we restrict ourselves to a precision of $10^{-3}$  for $\rho_c\sigma^3$ and $10^{-5}$ for $\beta_c\epsilon_c$ in the hard-sphere case and of $10^{-3}$ for $T_c/v_0$ and $10^{-5}$ for $\epsilon_c/v_0$ in the soft-sphere case.

\begin{figure}[t]
\begin{center}
\includegraphics[width=\columnwidth]{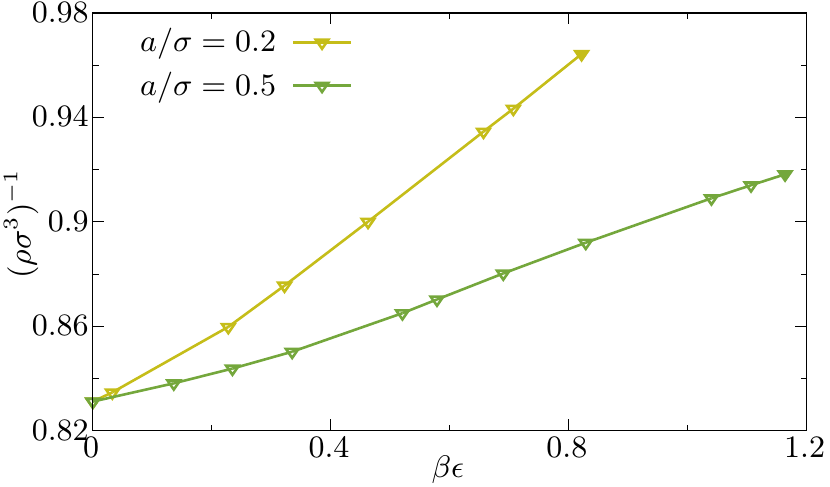} 
\caption{Phase diagram in the ($\beta\epsilon,~(\rho\sigma^3)^{-1}$) plane of the three-dimensional hard-sphere system in the HNC approximation. Two different values of the cutoff parameter $a$ are shown: $a/\sigma=0.2$ and $a/\sigma=0.5$. A line of first-order transition (empty symbols) emerges from the thermodynamic glass transition point in $\epsilon=0$ and ends in the critical point (full symbol) in ($\beta_c\epsilon_c,~(\rho_c\sigma^3)^{-1}$). Note the difference in the location of the line for the two values of $a$, except for the initial point in $\epsilon=0$, which represents the Kauzmann transition of the bulk system.}
\label{Fig_phase-diagram}
\end{center}
\end{figure}

We illustrate in Fig.~\ref{Fig_FPpotential} the behavior of the FP potential $V_a(\Delta Q)$ for the hard-sphere system as density increases for two different values of the cutoff parameter, $a/\sigma=0.2$ and $0.5$. The potential has a similar shape and evolution as first found in Refs. [\onlinecite{cardenas1999constrained},\onlinecite{cardenas1998glass}] (in their case $a/\sigma=0.3$). At $\rho_K \sigma^3=1.203$ the potential has two minima of equal height and the high-overlap minimum becomes metastable as $\rho$ decreases until it disappears in a saddle point at $\rho_d \sigma^3=1.183$ (above the value of $1.17$ found by in [\onlinecite{cardenas1999constrained},\onlinecite{cardenas1998glass}] but consistent with the value provided by Parisi and Zamponi\cite{parisi2010mean}). At still lower density the potential retains a nonconvex shape down to some critical density $\rho_c$ at which convexity is eventually recovered. As we have already emphasized, the values of $\rho_K$ and $\rho_d$ do not depend on the choice of $a$ but those of the overlap at the metastable minimum do depend on $a$. We also find, as will be further described below, that the value of the critical density $\rho_c$ depends on $a$ significantly.

\begin{figure}[!t]
\centering
\includegraphics[width=\columnwidth]{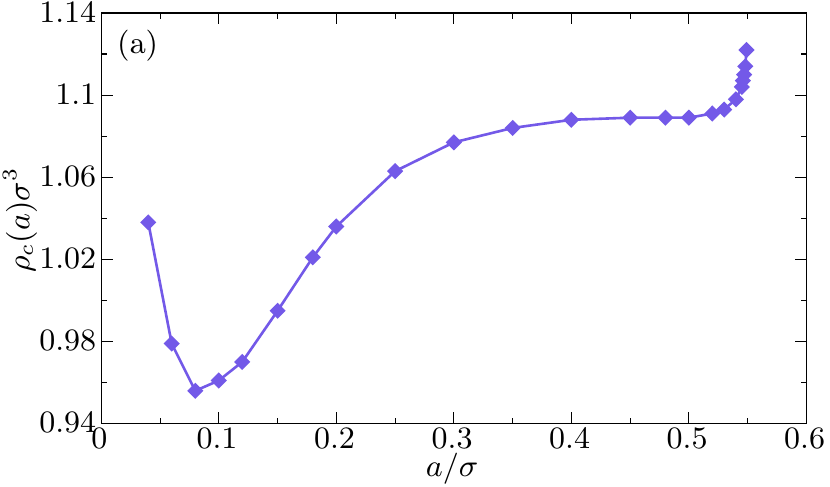} 
\includegraphics[width=\columnwidth]{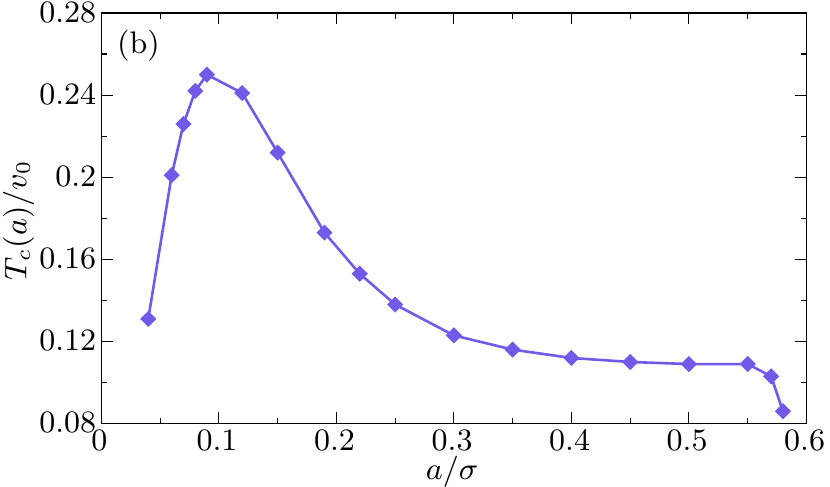} 
\caption{Variation with the cutoff parameter $a$ of the  critical density $\rho_c(a)$ for hard spheres (top panel) and of the critical temperature $T_c(a)$ for soft spheres (bottom panel) in the HNC approximation.}
\label{Fig_rho_c-T_c}
\end{figure}

In Fig.~\ref{Fig_phase-diagram} we display the phase diagram of the hard-sphere model in the ($\beta\epsilon,~(\rho\sigma^3)^{-1}$) plane for the same two values of $a$ as in Fig.~\ref{Fig_FPpotential}. As is well known\cite{franz1995recipes,franz1998effective,franz1997phase}, the nonconvexity of the FP potential gives rise to a line of first-order transition emerging from the thermodynamic glass transition point in $\epsilon=0$. The line ends in a critical point at ($\beta_c\epsilon_c,~(\rho_c\sigma^3)^{-1}$). As clearly seen, the location of the line is different for the two values of $a$, and the end critical point as well.  

\section{HNC results for the critical endpoint}

\label{sec:HNC_results}

\subsection{Numerical results}

In this section we systematically investigate the dependence on $a$ of the critical point that is associated with the return to convexity of the FP potential \revminor{in the HNC framework}. The critical density $\rho_c(a)$ for hard spheres or the critical temperature $T_c(a)$ for soft spheres is determined by solving the HNC equations, then using Eqs.~(\ref{eq_expressionDeltaQ}) and (\ref{eq_expression_FPpotential}) and the two conditions in Eq.~(\ref{eq_critical1}). Finally, $(\beta_c\epsilon_c)(a)$ is obtained from Eq.~(\ref{eq_critical2}).

\begin{figure}[t]
\centering
\includegraphics[width=\columnwidth]{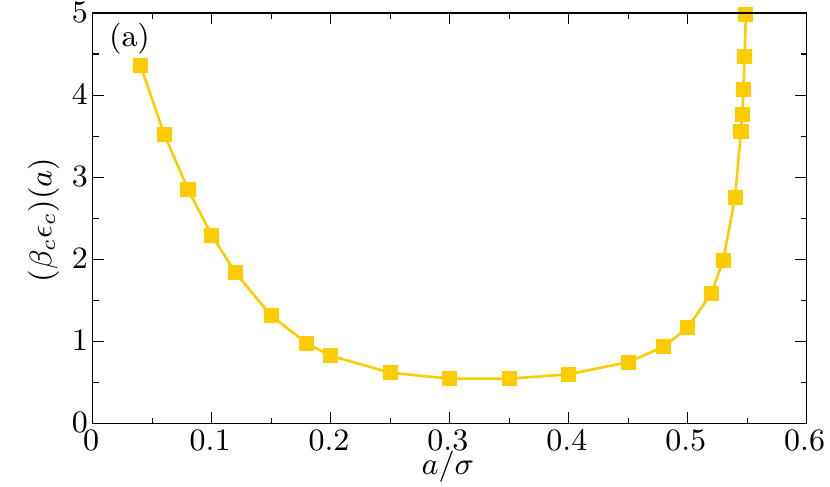} 
\includegraphics[width=\columnwidth]{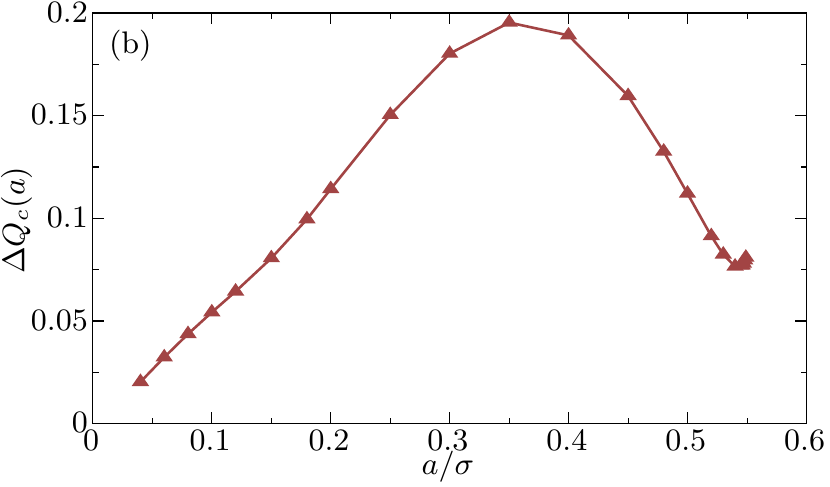} 
\caption{Variation with the cutoff parameter $a$ of the critical value of the source $(\beta_c\epsilon_c)(a)$ (a) and of the critical value of the overlap difference $\Delta Q_c(a)$ (b) for the three-dimensional hard sphere system in the HNC approximation.}
\label{Fig_critical-HS}
\end{figure}

We show in Fig.~\ref{Fig_rho_c-T_c} the variation with $a$ of the critical density $\rho_c(a)$ for hard spheres and the critical temperature $T_c(a)$ for soft spheres. Both critical quantities vary by a large amount: more than $15\%$ for $\rho_c$ and a factor of $2$ for $T_c$ over the covered range of $a$. For comparison, recall that within HNC the relative change between $\rho_d$ and $\rho_K$ for hard spheres is $1.7\%$ and between $T_d$ and $T_K$ for soft spheres is about $14\%$\footnote{For our model, $T_d/v_0=0.0535$ and $T_K/v_0=0.0464$.}. Furthermore, the evolution of either $\rho_c$ or $T_c$ with $a$ is nonmonotonic with a minimum in $\rho_c$ for $a\approx 0.08\sigma$ and a maximum in $T_c$ for $a\approx 0.09\sigma$. By choosing $a/\sigma$ around $0.08-0.09$ one can then move the critical point in the liquid phase quite significantly away from the dynamic and thermodynamic glass transitions, as compared with the conventional choice of $a=0.3\sigma$.

\begin{figure}[!t]
\centering
\includegraphics[width=\columnwidth]{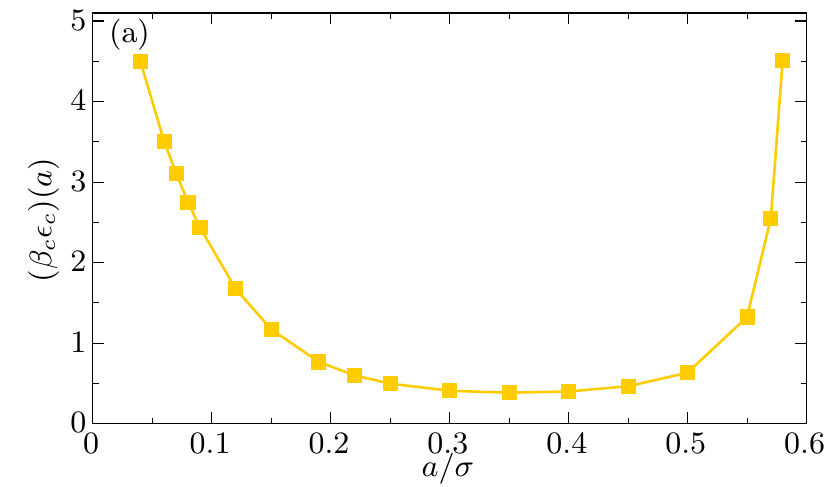} 
\includegraphics[width=\columnwidth]{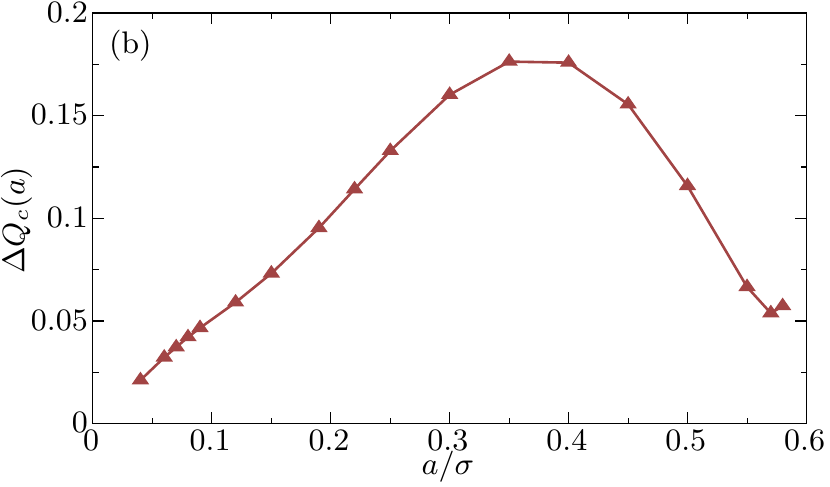} 
\caption{Variation with the cutoff parameter $a$ of the critical value of the source $(\beta_c\epsilon_c)(a)$ (a) and of the critical value of the overlap difference $\Delta Q_c(a)$ (b) for the three-dimensional soft sphere system in the HNC approximation.}
\label{Fig_critical-SS}
\end{figure}

The values of the source or coupling $\epsilon_c(a)$ and of the overlap $Q_c(a)$ (or rather of the difference $\Delta Q_c(a)$ with the random value) at the critical point are shown as a function of $a$ in Fig.~\ref{Fig_critical-HS} for the hard-sphere system and in Fig.~\ref{Fig_critical-SS} for the soft-sphere system. In all cases the variations with $a$ are nonmonotonic, with a minimum in $\beta_c\epsilon_c$ and a maximum in $\Delta Q_c$ around $a\approx 0.35\sigma$. The behavior of these critical quantities for vanishing and large values of $a$ will be discussed below.

\begin{figure}[!ht]
\centering
\includegraphics[width=\columnwidth]{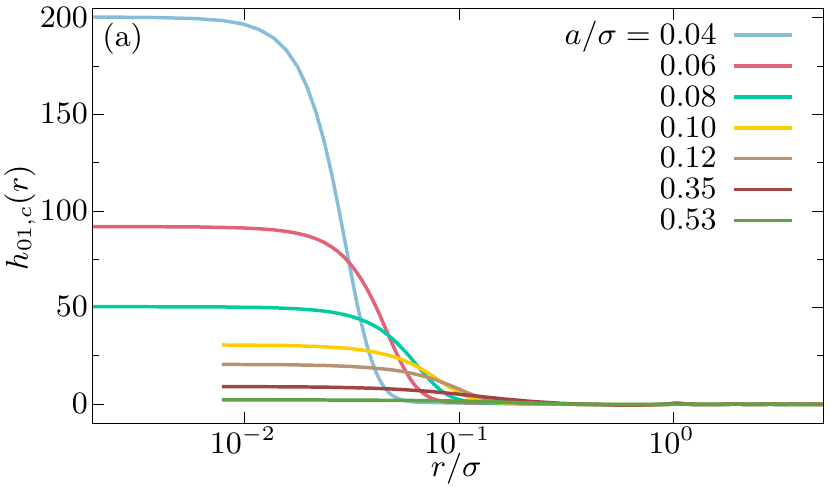} 
\includegraphics[width=\columnwidth]{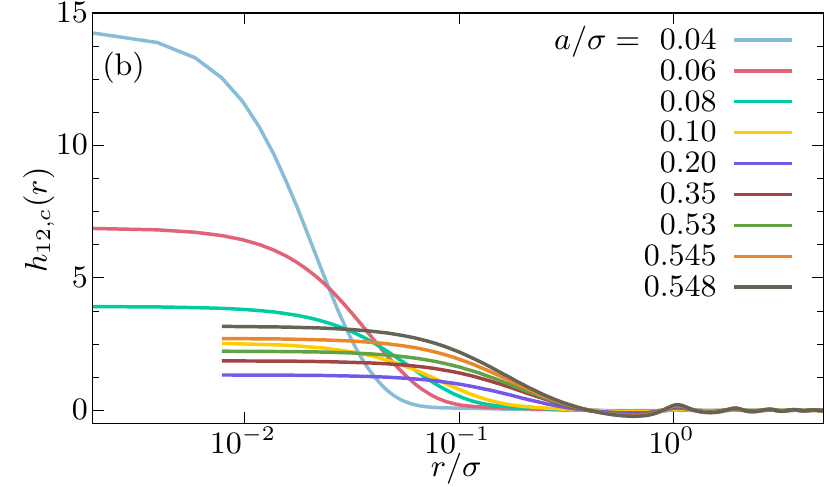} 
\includegraphics[width=\columnwidth]{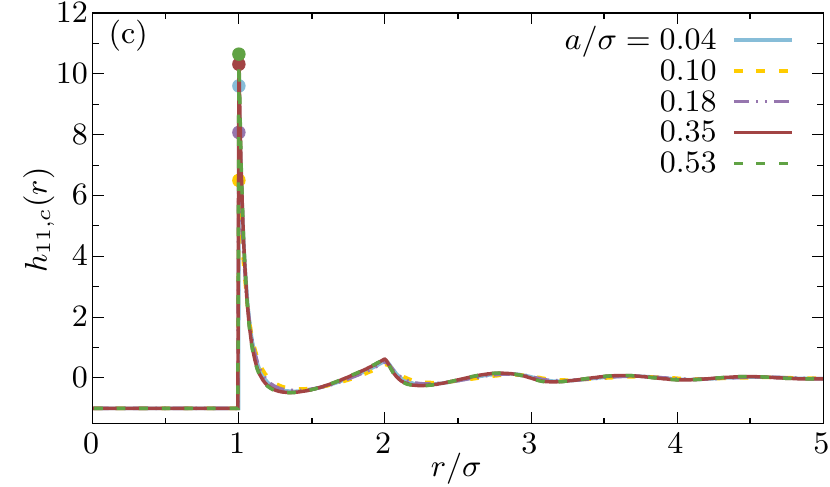}  
\caption{HNC total correlation functions $h_{01,c}(r)$ (a), $h_{12,c}(r)$ (b), and $h_{11,c}(r)$ (c) versus $r/\sigma$ (on a logarithmic scale for the two first panels and on a linear scale for the last one) at criticality for a wide range of values of $a$ for hard spheres in the HNC approximation. In panel (c) the dot marks the maximum value of $h_{11,c}$.}
\label{Fig_critical_h}
\end{figure}

\rev{Note that the variation with $a$ of the location of the critical point is not given by a simple dimensional analysis, $\Delta Q_c(a)\sim a^3$ and $(\beta_c \epsilon_c)(a)\sim 1/a^3$ (in the framework of the replicated free-energy functional presented in Sec.~\ref{sec:statmech}, $\epsilon$ and $a$ represent the strength and the range of the attractive interaction between replicas). The observed nonmonotonic behavior and the detailed evolution at small $a$ and large $a$ are much more involved than this naive expectation. This stems from the nontrivial structure and variation with $a$ of the pair correlation functions.}

We display in Fig.~\ref{Fig_critical_h} the HNC total correlation functions $h_{01,c}(r)$, $h_{12,c}(r)$, and $h_{11,c}(r)$ at criticality for a wide range of values of $a$ in the case of the hard-sphere system.  Note that due to the hard-core exclusion $h_{11,c}(r)=-1$ for $r<\sigma$. On the other hand, $h_{01,c}(r)$ and $h_{12,c}(r)$ have a nontrivial $r$ dependence on a scale $r\sim a<\sigma$, and their value at small $r \ll a$ strongly increases as $a$ decreases when $a\leq 0.2\sigma$. We discuss this behavior in the next section.

\subsection{Behavior at small values of $a$}

We consider first the limit in which $a\to 0^+$, where as seen from Figs.~\ref{Fig_critical-HS} and \ref{Fig_critical-SS}, $\Delta Q_c(a)$ seems to go to $0$ whereas $\left(\beta_c\epsilon_c\right)(a)$ seems to diverge. To make some progress in trying to rationalize this limiting behavior, we assume that $\rho_c(a)$ and $T_c(a)$ stay finite and nonzero when $a\to 0^+$, which is compatible with the data in Fig.~\ref{Fig_rho_c-T_c}, and that the total correlation functions $h_{01,c}(r)$ and  $h_{12,c}(r)$ can be decomposed in a part that varies on the scale of $a$, whose amplitude grows as $a\to 0^+$, and a part that varies on the scale of $\sigma$, whose amplitude goes to zero as $a\to 0^+$. (Note that when $a=0$, the replicas are decoupled, $h_{12}=h_{01}\equiv 0$ and $h_{11}=h_{00}$.) As already noticed, the function $h_{11,c}(r)$ on the other hand only varies on the scale of $\sigma$ with a $O(1)$ amplitude, and so does $h_{00,c}(r)$ (which is independent of $a$). 

Through heuristic arguments based on an analysis of the HNC equations in the limit $a\to 0^+$ we derive that a consistent solution of the equations is obtained for the total and direct correlation functions at criticality in the form (for convenience we omit the subscript $c$ on all the quantities)
\begin{equation}
\begin{aligned}
 \label{eq_small-a_h}
&h_{01}(r)= a^{-3/2}|\ln a|^{1/2}\hat h_{01}(r/a)+a^{3/2}|\ln a|^{1/2} \tilde h_{01}(r/\sigma)\\&
h_{12}(r)= a^{-3/2}|\ln a|^{1/2}\hat h_{12}(r/a)+a^{3/2}|\ln a|^{1/2} \tilde h_{12}(r/\sigma)\\&
h_{11}(r)=\tilde h_{00}(r/\sigma) + {\rm O}(a^{3}|\ln a|)\\&
h_{00}(r)=\tilde h_{00}(r/\sigma)\, ,
\end{aligned}
\end{equation}
where all the functions  $\hat h_{\alpha\gamma}(x)$ and $\tilde h_{\alpha\gamma}(x)$ have an amplitude and a range of $O(1)$. The function $h_{{\rm con}}$ is the difference between $h_{11}$ and $h_{12}$ given by the above expressions.

\begin{figure}[tb]
\begin{center}
\includegraphics[width=\columnwidth]{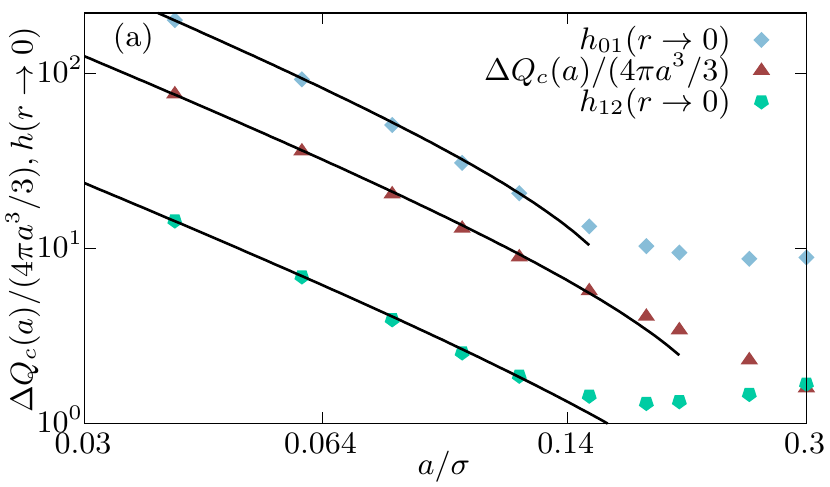} 
\includegraphics[width=\columnwidth]{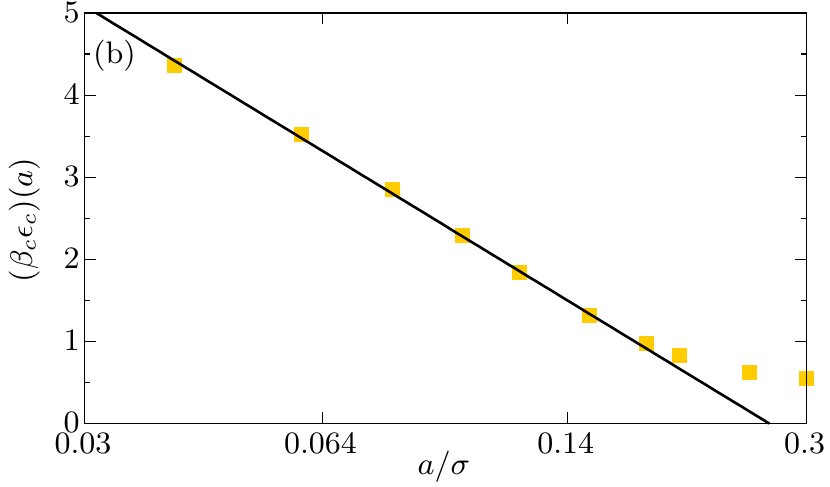} 
\caption{Limit $a\to 0^+$ of the HNC calculation in the case of hard spheres. (a) Log-log plot of $\Delta Q_c/(4\pi a^3/3)$, $h_{01}(r\to 0)$ and $h_{12}(r\to 0)$ versus $a/\sigma$; the continuous lines show the expected $a^{-3/2}|\ln a|^{1/2}$ dependence. (b) $\beta_c\epsilon_c$ on a linear scale versus $a/\sigma$ on a logarithmic scale along with the expected $\vert \ln a\vert$ behavior (continuous line).}
\label{Fig_HNC_small-a}
\end{center}
\end{figure}

In Fourier space, the above expressions translate into
\begin{equation}
\begin{aligned}
 \label{eq_small-a_Fourier_h}
&h_{01}(q)= a^{3/2}|\ln a|^{1/2}\left[\hat h_{01}(qa)+\sigma^{3}\tilde h_{01}(q\sigma)\right]\\&
h_{12}(q)= a^{3/2}|\ln a|^{1/2}\left[\hat h_{12}(qa)+\sigma^{3} \tilde h_{12}(q\sigma)\right]\\&
h_{11}(q)= \sigma^{3}\tilde h_{00}(q\sigma) + {\rm O}(a^{3}|\ln a|)\\&
h_{00}(q)= \sigma^{3}\tilde h_{00}(q\sigma) ,
\end{aligned}
\end{equation}
where we have kept the same notation for the functions in real and Fourier spaces. Note that both $h_{01}(q)$ and $h_{12}(q)$ go to $0$ when $a\to 0$. The tilde functions keep the signature of the liquid structure and have a peak near $q\approx 2\pi/\sigma$ whereas the hat functions have a structure that follows from that of $w$ and decay on a range $q\approx 1/a$. This implies that a complete separation of scales for the wave-vector dependence of the tilde and hat functions is achieved when $2\pi/\sigma \ll 1/a$; this requires in practice very small values of $a$, typically, $a/\sigma\lesssim 10^{-2}$. Details on the derivation are given in Appendix~\ref{app_analysisHNC}.

With the above ansatz, one has
\begin{equation}
\begin{aligned}
 \label{eq_small-a}
&\Delta Q_c(a\to 0^+)\sim a^{3/2}|\ln a|^{1/2}4\pi\rho_c(0^+)\int_0^\infty \!\!\!\!dx x^2 w(x) \hat h_{01}(x)\\&
\left(\beta_c\epsilon_c\right)(a\to 0^+)\sim \widehat{\beta\epsilon}\vert \ln a\vert,
\end{aligned}
\end{equation}
so that $\beta_c\epsilon_c \Delta Q_c\to 0$ as $(a|\ln a|)^{3/2}$ when $a\to 0^+$. 

We compare the above predictions with the numerical solution of the HNC equations for small $a$ in Fig.~\ref{Fig_HNC_small-a}. One can check that $\Delta Q_c/(4\pi a^3/3)$,  $h_{01}(r\to 0)$, and $h_{12}(r\to 0)$ all diverge as $a^{-3/2}|\ln a|^{1/2}$ [panel (a)] and that $\beta_c\epsilon_c$ diverges as $|\ln a|$ [panel(b)], as expected from the above equations. Additional comparisons between numerical results and analytical predictions are provided in Appendix~\ref{app_analysisHNC}.

\subsection{Behavior at large values of $a$}

Finally, we discuss the case of large values of $a$. As can be seen from the bottom panels of Figs.~\ref{Fig_critical-HS} and \ref{Fig_critical-SS}, the overlap difference with the random value of the overlap (which gives the location of the stable liquid minimum of the FP potential) $\Delta Q_c(a)$ decreases as $a$ increases for $a/\sigma \gtrsim 0.35$ and seems to stick to a finite value for $a/\sigma\approx 0.55$. For $a\gtrsim 0.55$, the numerical solutions of Eqs.~(\ref{eq_HNC_Fourier}) and (\ref{eq_HNC_closure}) become more difficult to follow even for $\rho\geq \rho_d$ (or $T\leq T_d$). At the same time, the HNC integral equations do not seem to be driven to any singularity.

\begin{figure}[!t]
\centering
\includegraphics[width=\columnwidth]{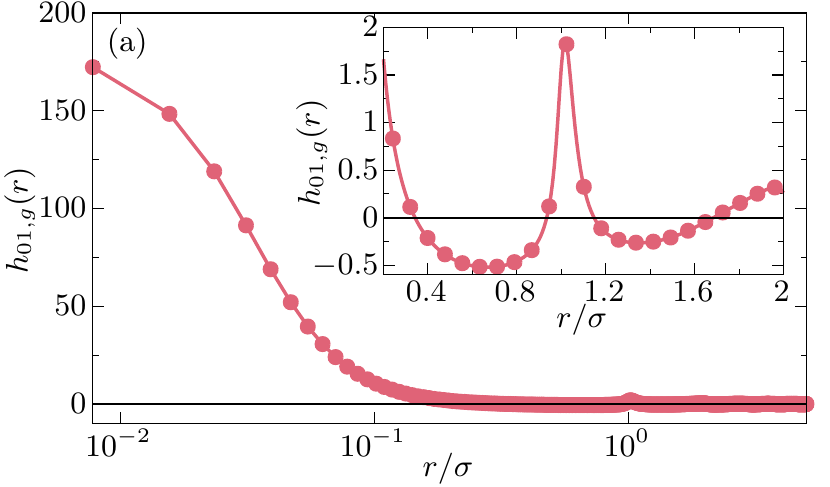} 
\includegraphics[width=\columnwidth]{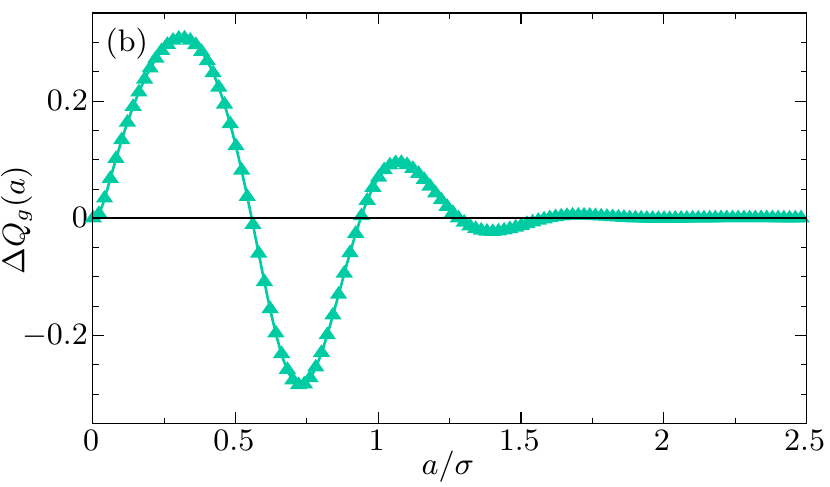} 
\caption{HNC result for the metastable glass minimum of the FP potential for hard spheres at a density $\rho=1.193$ which is intermediate between $\rho_d$ and $\rho_K$: (a) total correlation function $h_{01,g}(r)$; (b) difference in overlap $\Delta Q_g$ with the global minimum as a function of $a$.}
\label{Fig_h-Q_glass}
\end{figure}

To try to understand this behavior, it is worth looking first at what happens at the metastable minimum when the latter exists beyond the dynamical transition. For concreteness we focus on the hard-sphere model. As we have already noted, the total correlation functions at the minima of the FP potential are independent of $a$. Let us call $\Delta Q_g(a)$ the difference between the overlap at the metastable glass minimum and that at the global minimum for $\rho\geq \rho_d$. Then, from Eq.~(\ref{eq_expressionDeltaQ}),
\begin{equation}
\label{eq_expressionDeltaQ_glass}
\Delta Q_g(a)= 4\pi \rho \int_{0}^\infty dr r^2 w(r/a) h_{01,g}(r)\,,
\end{equation}
with $h_{01,g}(r)$ independent of $a$. Because $h_{01,g}(r)$ becomes negative for $r \gtrsim 0.35\sigma$ [see for illustration the function at a density $\rho_d<\rho<\rho_K$ in Fig.~\ref{Fig_h-Q_glass}(a)], the integral in Eq.~(\ref{eq_expressionDeltaQ_glass}) can become negative for some values of $a$. This is shown in Fig.~\ref{Fig_h-Q_glass}(b) where we plot $\Delta Q_g$ as a function of $a$: it is positive for small values, then turns negative for $a/\sigma\geq0.556$, becomes positive again for $a/\sigma \geq0.938$ and eventually weakly oscillates around a slightly positive value. This is found for all densities above $\rho_d$, and in the ideal glass phase as well. The value $a_*/\sigma$ for which $\Delta Q_g(a)$ first turns negative does not vary much with density (it is equal to $0.5576$ at $\rho_d$ and $0.555$ at $\rho_K$). So, while the underlying physics is unchanged, by changing the cutoff parameter in the definition of the overlap, one can switch from correlated replicas at the metastable glass minimum ($\Delta Q_g>0$) to anti-correlated replicas ($\Delta Q_g<0$). For physical reasons, it seems more pleasant to work with $\Delta Q_g>0$ and restrict the range of $a$ to $a<a_*$, but this restriction is not motivated by the presence of a physical singularity.

From the above considerations, one can rationalize the behavior of the critical point as $a/\sigma$ approaches some special value close to $0.55$. Replacing for simplicity the smooth $w(r/a)$ by a discontinuous step function, one finds that $\Delta Q_c(a)\approx 4\pi\rho_c(a)\int_0^a dr r^2 h_{01,c}(r)$. The maximum observed in $\Delta Q_c(a)$ should then appear in the close vicinity of the value of $a$ for which $a=r_*(a)$, where $r_*(a)$ is the lowest $r$ for which $h_{01,c}(r)=0$. This is indeed what is numerically found with $a_{{\rm max}}/\sigma\approx 0.35$ while the value of $a$ such that $a=r_*(a)$ is $a\approx 0.39\sigma$. For $a>a_\mathrm{max}$, $\Delta Q_c(a)$ decreases because the integral involves negative values of $h_{01,c}(r)$. Therefore, when $\Delta Q_c$ becomes too small, all nontrivial features of the FP potential become concentrated essentially in a point and one can no longer numerically solve Eqs.~(\ref{eq_critical1}) and (\ref{eq_critical2}). Again, this is not associated with any physical phenomenon. Except for a small region  $0.555\leq a \leq 0.5576$ (see above) where an unrealistic behavior of the phase transition line between low-overlap and high-overlap phases is found (a peculiarity  that does not seem worth studying in more depth), larger values of $a$ (but still lower than the next value of $a$ for which $\Delta Q_g$ vanishes) correspond to a well-behaved first-order transition line, yet with a critical endpoint characterized by $\Delta Q_c<0$ and $\beta_c\epsilon_c<0$.


\begin{figure}[!t]
\centering
\includegraphics[width=\columnwidth]{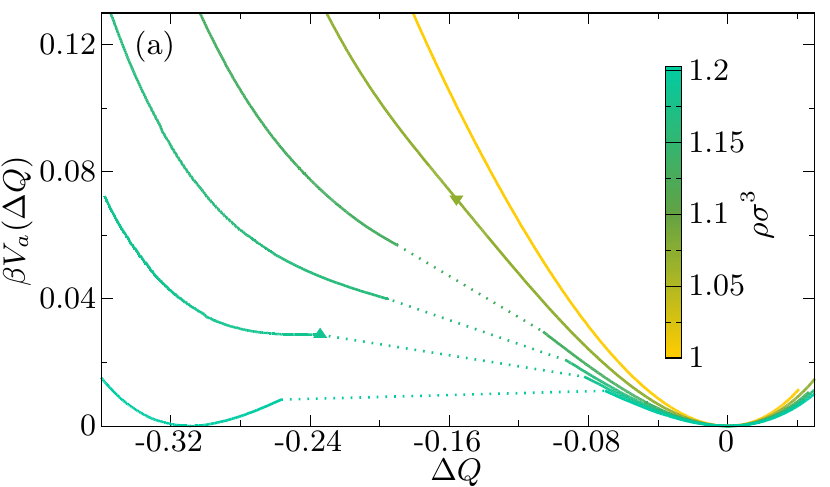} 
\includegraphics[width=\columnwidth]{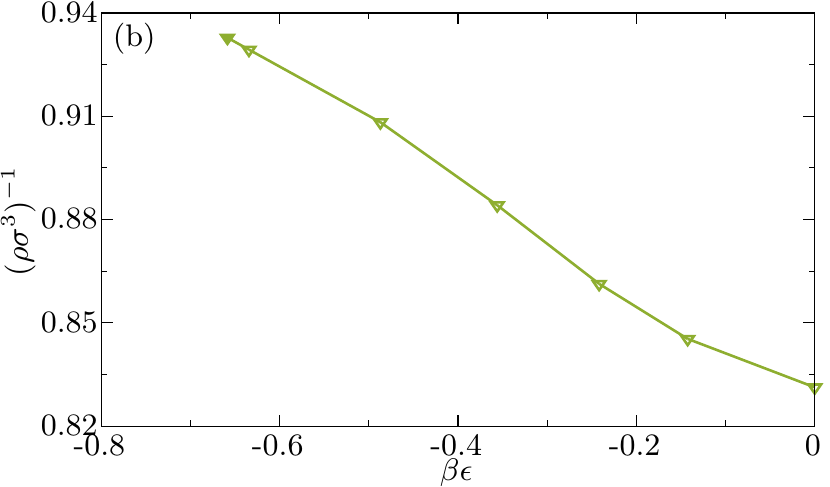} 
\caption{HNC result for three-dimensional hard spheres and $a/\sigma=0.73$: (a) FP potential (the dotted lines represent the region where there is no replica-symmetric solution to the HNC equations\rev{\cite{Note5}}); (b) phase diagram in the ($\beta\epsilon,~(\rho\sigma^3)^{-1}$) plane (the critical point is emphasized with a closed symbol).}
\label{Fig_diagram-large-a}
\end{figure}
We illustrate this feature for a value of the cutoff parameter $a=0.73\sigma$. In Fig.~\ref{Fig_diagram-large-a}(a), we plot the FP potential, which has the same behavior as in Fig.~\ref{Fig_FPpotential} except that all its noticeable characteristics are located in the range $\Delta Q<0$. In particular, a critical point is indeed found with $\beta_c\epsilon_c<0$ and $\Delta Q_c<0$, as illustrated in Fig.~\ref{Fig_diagram-large-a}(b).

\section{Computer simulations}
\label{sec:simulation}

\begin{figure}[t]
\includegraphics[width=\columnwidth]{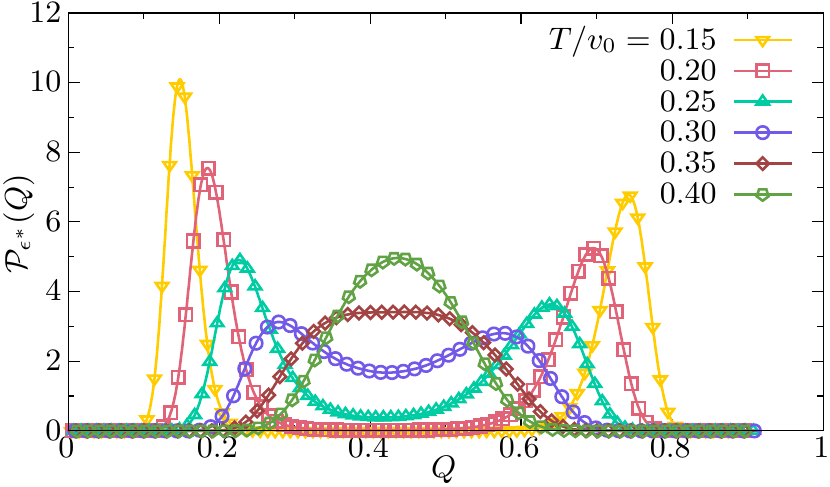}
\caption{\rev{Probability distribution of the overlap $\mathcal{P}_{\epsilon^*}(Q)$ when a source $\epsilon^*(T)$ linearly coupled to the overlap is applied. The field $\epsilon^*(T)$ corresponds to the locus of the maximum of total susceptibility $\chi_\mathrm{con}+\chi_\mathrm{dis}$ at fixed temperature. These distributions were obtained from computer simulations of a polydisperse mixture of $N=600$ soft spheres with $a=0.22\sigma$ using umbrella sampling and reweighting. For this value of $a$, the critical point is located at $T_c\approx 0.167v_0$\cite{guiselin2020random}.}}
\label{Fig_pdf}
\end{figure}

To complement the detailed but approximate analysis obtained through the HNC treatment we have studied a three-dimensional glass-forming liquid model of soft spheres by computer simulation, in which we rely on the recently developed swap algorithm\cite{berthier2016equilibrium,ninarello2017models}. We consider a polydisperse mixture of spherical particles of diameters $\sigma_i$ distributed according to the distribution $p(\sigma_i)\propto \sigma_i^{-3}$ for $\sigma_i\in\left[\sigma_\mathrm{min};\sigma_\mathrm{max}\right]$, with $\sigma_\mathrm{min}=0.726\sigma$ and $\sigma_\mathrm{max}=1.6095\sigma$ where $\sigma$ is the average diameter, as in Refs.~[\onlinecite{berthier2017configurational},\onlinecite{berthier2019efficient}]. In addition, the interaction potential has the same analytical form as in soft-sphere model studied in the above HNC treatment, but the cross-diameters $\sigma_{ij}$ are nonadditive to prevent crystallization and demixing\cite{ninarello2017models}: $\sigma_{ij}=0.5\left(\sigma_i+\sigma_j\right)\left(1-0.2\left\vert\sigma_i-\sigma_j\right\vert\right)$. We have already studied in detail the critical endpoint of this liquid\cite{guiselin2020random} with the specific choice $a=0.22\sigma$; most of the simulations were done when the temperature $T_0$ of the reference replica $0$ is different from the temperature $T$ of the constrained replicas and fixed to a low value $T_0=0.06v_0\gtrsim T_g$ (with $T_g$ the estimated laboratory glass-transition temperature). 

\rev{In Ref.~[\onlinecite{guiselin2020random}], by using extensive computer simulations, we have studied the size and temperature dependences of the so-called ``connected susceptibility'' $\chi_\mathrm{con}=N\beta\left[\overline{\langle Q^2\rangle-\langle Q\rangle^2}\right]$ (with $\langle \cdots\rangle$ denoting the thermal average and $\overline{\vphantom{Q}\cdots\vphantom{Q}}$ the average over the quenched disorder represented by the reference configuration), and of the so-called ``disconnected susceptibility'' $\chi_\mathrm{dis}=N\beta\left[\overline{\langle Q\rangle^2}-\overline{\langle Q\rangle}^2\right]$, as usual for random-field-like systems. The former quantifies thermal fluctuations, the latter disorder-induced fluctuations. Both are expected to diverge at the critical point in the thermodynamic limit, when $\epsilon$ goes to $\epsilon_c$ and $T$ to $T_c$. In finite-size systems, the susceptibilities should instead behave as power laws of the linear size of the system, with the exponents characterizing the universality class of the critical point. In [\onlinecite{guiselin2020random}], we have performed a finite-size scaling analysis and we have shown that when temperature and susceptibilities are properly rescaled with system-size dependent prefactors and the known critical exponents of the three-dimensional random-field Ising model (RFIM), data from different sizes and temperatures all collapse on a single master-curve. This shows that the critical point survives in a $3$-dimensional glass-forming liquid, in spite of the presence of finite-dimensional fluctuations on all scales, and belongs to the universality class of the RFIM.} Here, we build on this study to investigate the influence of the cutoff parameter $a$ on the position of the critical point in the $\left(\epsilon,~T\right)$ phase diagram.

\rev{For computational efficiency (these studies are highly demanding in terms of computer time), instead of looking at the case where the reference and the constrained liquid configurations are at the same temperature $T$, we focus on the situation where the temperature $T_0$ is fixed to a low value $0.06v_0$ (at which reference configurations can nonetheless be equilibrated thanks to the swap algorithm). Then, the critical endpoint moves up in temperature compared to the situation $T_0=T$, for all values of $a$, which results in a considerable speedup of the simulations\cite{franz1998effective}.} To allow for a comparison with HNC predictions, we have repeated the HNC treatment for the case where $T_0\neq T$ and the single-component soft-sphere liquid. To be quantitatively similar with the choice in the simulations, we have chosen a $T_0$ intermediate between $T_d$ and $T_K$ and then solved the equations of the HNC approximation. 

\begin{figure}[!ht]
\includegraphics[width=\columnwidth]{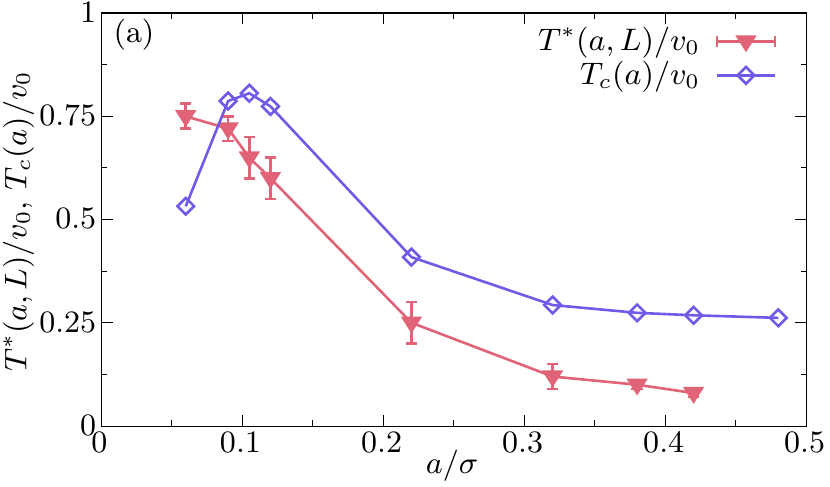}
\includegraphics[width=\columnwidth]{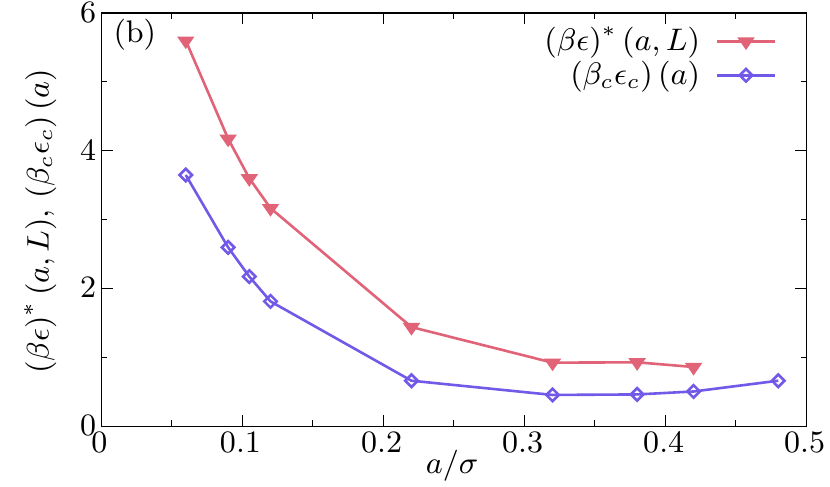} 
\includegraphics[width=\columnwidth]{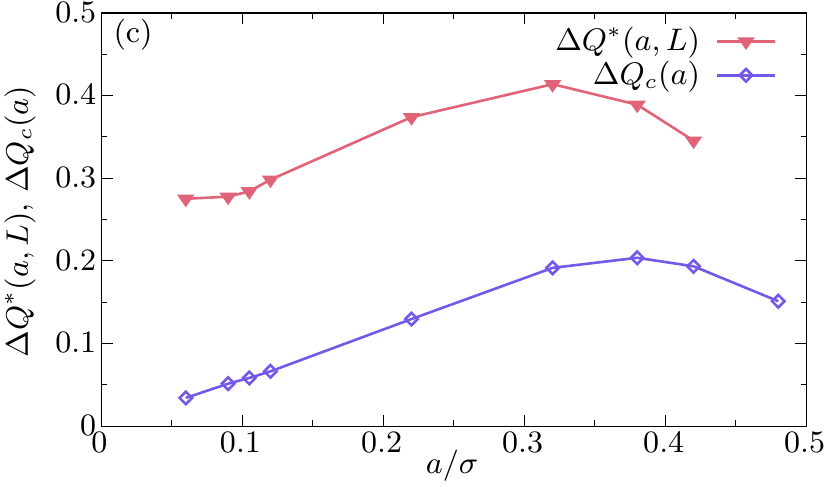} 
\caption{Comparison between the results of a computer simulation of a polydisperse mixture of $N=L^3=600$ soft spheres with a reference configuration at a low temperature $T_0=0.06 v_0$ and the HNC calculation of a single-component soft-sphere liquid with a reference configuration at temperature $T_K<T_0=0.0499v_0<T_d$. (a) Location of the maximum of the low-overlap connected susceptibility $T^*(a,L)$ for the simulation and of the critical temperature $T_c(a)$ for the HNC calculation. \rev{(b)-(c) Estimate of the critical source $\left(\beta_c\epsilon_c\right)(a)$ and of the overlap difference with the random limit at criticality $\Delta Q_c(a)$ from the simulation and the HNC approximation.}}
\label{Fig_critical_T_simu}
\end{figure}

The statistical properties of the overlap are computed in the simulations thanks to umbrella sampling and a subsequent reweighting\cite{guiselin2020random}. \rev{This strategy enables us to compute the Franz-Parisi potential but also all the thermodynamic properties of the liquid when coupled to the quenched reference with an arbitrary applied source $\epsilon$. In particular, for each temperature of the simulation, we can find the probability distribution $\mathcal{P}_\epsilon(Q)$ of the overlap for the value of the source $\epsilon=\epsilon^*(T)$ that maximizes the total susceptibility, defined as the sum of $\chi_\mathrm{con}$ and $\chi_\mathrm{dis}$. (This defines the analog of the ``Widom line'' above the standard gas-liquid critical point.) These distributions are shown in Fig.~\ref{Fig_pdf} for a system of size $N=600$ particles with $a=0.22\sigma$. In Ref.~[\onlinecite{guiselin2020random}], we have estimated that for the same value of $a$, $T_c\approx 0.167$. Consequently, the probability distribution of the overlap becomes bimodal for temperatures significantly above the critical temperature $T_c$, and at these temperatures one can study overlap fluctuations  restricted either to the low-overlap peak or to the high-overlap peak. In the following, we focus on the connected susceptibility in the low-overlap phase $\chi_\mathrm{con}^{\mathrm{low}}$ measured at $\epsilon^*(T)$ for each temperature.}

Rigorously, as already mentioned, the location of the critical point in a simulation study can only be found through a finite-size scaling analysis. For instance, taking into account the $a$-dependence, the low-overlap connected susceptibility $\chi_\mathrm{con}^{\mathrm{low}}$ should scale as $\chi_\mathrm{con}^{\mathrm{low}}(T,L,a)=B_aL^{\gamma/\nu}\tilde\chi(y_atL^{1/\nu})$, with $t=T/T_c(a)-1$ the reduced temperature, $L\propto N^{1/3}$ the linear size of the system, $\gamma$ and $\nu$ the critical exponents of the $3d$-RFIM, $\tilde{\chi}(x)$ a universal scaling function, and $B_a$ and $y_a$ $a$-dependent constants. The scaling function has a maximum 
for $x=x^*$, which corresponds to a temperature $T^*(a,L)=T_c(a)\left(1+x^*L^{-1/\nu}/y_a\right)$. Assuming that $y_a$ depends only slightly on $a$, the measure of $T^*(a,L)$ at fixed system size gives a reasonable proxy for the evolution of the critical temperature with $a$ in this system (but the absolute value of the temperature itself is still too high). 


A comparison between the results of the simulation and the HNC ones is shown in Fig.~\ref{Fig_critical_T_simu}. The trends as $a$ decreases are very similar. The HNC prediction for the critical temperature $T_c(a)$ passes through a maximum around $a/\sigma\approx 0.09$ whereas the simulation data appear to plateau at the lowest studied values. \rev{(It is unclear if this difference would persist at even lower values of $a$ in the simulation or with a better determination of the critical temperature; studying such small values of $a$ however becomes computationally prohibitively costly.) The agreement is also good when comparing the evolution of the critical value of the source $\beta \epsilon$: for both simulation and HNC results, this quantity first decreases with increasing $a$. The HNC prediction for $(\beta_c\epsilon_c)$ reaches a minimum for $a/\sigma\approx 0.35$ and subsequently increases with $a$ slowly while it seems to plateau in the simulations. However, for the latter, the critical point falls in a temperature range for which equilibration becomes difficult to ensure, even with the swap algorithm. Consequently, we cannot state whether the quantity $(\beta_c\epsilon_c)$ would eventually increase when $a$ gets even larger (or whether this tendancy would remain with a better determination of the critical point). The evolution of $\Delta Q_c(a)$ strengthens the agreement between the simulation and the HNC calculations: in both cases, we observe an increase with $a$ at small values of $a$, followed by a maximum for $a/\sigma\approx 0.35$, and a subsequent decrease. All in all, and in spite of expected discrepancies due to the difference in polydispersity, a possible dependence on $a$ of the finite-size effects in the simulations, and the absence of nontrivial long-range fluctuations in the HNC calculations, the evolutions with $a$ seen in the simulation and in the HNC treatment of a $3$-dimensional glass-forming liquid are thus in qualitative agreement.}

\section{Conclusion}
\label{sec:conclusions}

The similarity or overlap between pairs of configurations has proven a powerful concept to describe the complex free-energy landscape of glassy systems and it furthermore provides the order parameter for the glass transition at the mean-field level. Whereas for lattice spin models the definition of the overlap is rather straightforward, it is somehow ambiguous in the case of glass-forming liquids. The overlap or similarity must then be defined up to some tolerance, typically a fraction $a/\sigma$ of the inter-particle distance. In this paper we have systematically investigated the dependence of the overlap fluctuations and of the phase diagram obtained by linearly coupling the overlap to an applied source on the parameter $a/\sigma$ in three-dimensional models of glass-forming liquids.

Within a general framework based on liquid-state theory and using for illustration the hypernetted-chain (HNC) approximation, we show that while the dynamical and thermodynamic glass transitions found in this mean-field-like approximation \rev{of a $3$-dimensional glass-forming liquid} are independent of $a/\sigma$, the whole extended phase diagram involving a transition between a low-overlap phase and a high-overlap one in the presence of an applied source (or coupling) strongly depends on the value of $a/\sigma$. In the theoretical framework, this can be understood by noting that the singular features of the underlying functional of the correlation functions (the so-called Morita-Hiroike functional) are independent of $a/\sigma$ but that the precise choice of the order parameter which requires fixing the value of $a/\sigma$ influences the phase diagram, except for the minima obtained in zero source. We are able to rationalize through analytical and numerical arguments the evolution of the location of the critical point (ending the transition line between low-overlap and high-overlap phases) for small and large values of $a/\sigma$ and we also confirm the theoretical predictions by computer simulations of a three-dimensional polydisperse glass-forming liquid.

\rev{At the level of the HNC approximation, we find in particular that the location of the terminal critical point obtained for a nonzero applied source follows a nonmonotonic behavior in temperature, density, or coupling strength as a function of $a/\sigma$. The most interesting feature for a practical application to computer simulations is that the critical temperature $T_c$ is pushed up by a factor of $2$ or more for values of $a/\sigma$ that are significantly lower than the values, $a/\sigma\approx 0.2-0.3$, systematically taken in previous studies involving overlaps in glass-forming liquids. The critical point then appears in the liquid region where viscosity is low and equilibration may be significantly faster. (A similar effect is found when density is the control parameter but the relative change is of course smaller although still of the order of $10\%$.) However, there are practical limitations to taking too small values of $a/\sigma$. In molecular-dynamics simulations, the magnitude of the forces exerted by the reference configuration when $\epsilon>0$ increases with decreasing $a$, forcing one to reduce the time-step in the integration of the equations of motion. On the other hand, in Monte-Carlo simulations, significant variations of the overlap are triggered by smaller amplitudes of the particle displacements as $a/\sigma$ is reduced, which requires trial moves of smaller size. The trade-off between shifting up the critical temperature and maximizing the algorithmic efficiency (simulated physical time versus computer time) therefore leads to an operational optimum value of $a$, which is around $0.1\sigma$. (This is valid whether a true maximum or a plateau exists in the curve $T_c(a)$ for small values of $a$.)}

\rev{Choosing $a$ around $0.1\sigma$ would then significantly accelerate computer simulations in the context of the study of the $(\epsilon,T)$ phase diagram, offering the opportunity to consider larger system sizes than considered so far. In addition, this choice could prove useful for glass-forming liquid models for which the swap algorithm is inefficient and cannot provide reference equilibrium configurations at a low temperature $T_0$ to shift up the critical temperature. Indeed, almost all previous simulation attempts~\cite{franz1998effective,cammarota2010phase,berthier2013overlap,berthier2015evidence} to study the critical point in $3$-$d$ model glass-formers, with the typical choice $a/\sigma\approx 0.2-0.3$ and both the reference and the constrained replicas at the same temperature, have been limited in practice to temperatures \textit{above} the putative critical temperature $T_c$ and to rather small system sizes. Indeed, $T_c$ seems to fall close to the mode-coupling crossover which represents the lowest temperature for which equilibration can be ensured in a reasonable computer wall-time without using the swap algorithm. By choosing an optimized value of $a$ for the definition of the overlap, one could more convincingly study the existence and the properties of the critical point in a variety of models of glass-forming liquids.}

\rev{Before concluding, let us comment on why the issue of the critical point at $T_c$ is an important one for the theory of the glass transition. As already stressed, the whole construction involving coupling equilibrium configurations of a liquid to a reference configuration of the same liquid is a tool to investigate the statistical properties of the liquid landscape in configurational space, thereby going beyond the description in terms of standard structural and thermodynamic quantities. While the construction seems difficult to reproduce in actual experiments on molecular glass-forming liquids, it can be implemented in computer simulations of glass-forming liquid models. One can then hope to assess whether the mean-field description of glass formation based on the complexity of the underlying free-energy landscape and the multitude of metastable states keep some relevance in $3$-dimensional glass-formers. To stay at the level of static properties (dynamics is not considered here), the Franz-Parisi potential is a quantity of choice to look for vestiges of the mean-field scenario, as illustrated in Fig. \ref{Fig_sketch}. The hypothetical glass transition at the Kauzmann temperature $T_K$ is unreachable and the spinodal/dynamical transition at $T_d$ is avoided, but the persistence of the mean-field scenario in finite dimensions requires the presence of the critical endpoint $T_c$ in the extended phase diagram: if there is no $T_c$, there is no $T_K$. Being able to more thoroughly investigate, through a large span of system sizes and a finite-size scaling analysis, the presence of a critical point in a variety of liquid models and check its universality class is therefore a worthwhile endeavor, which the outcome of the present study should facilitate.}

\rev{To conclude, we address the physical meaning, if any, of the dependence that we have found on the tolerance parameter $a$ involved in the definition of the similarity between liquid configurations. As the theoretical framework aims at unfolding key properties of the rather abstract liquid configurational space, rather than studying structural and thermodynamic indicators that are directly accessible to experiments, it is hard to provide a simple interpretation, even more so beyond the mean-field setting. The advantage of the overlap order parameter is to define some sort of metric between configurations that allows one to sort them in ``metastable states'', when the liquid is in equilibrium at a temperature $T$. It is expected that the distinguishing property of such states is not their free-energy density and that their number is an important factor.  But, how exactly similar should two liquid configurations be to be considered as belonging to the same state in the complex landscape? Varying the tolerance $a$ is a way to check how the properties of the coarse-grained landscape depend on the more or less strict definition of the similarity, hence on the coarse-graining length. (Note that this question is connected to, but is different from, the issue of the lifetime of a metastable state, which must be finite in a finite-dimensional liquid and therefore imposes a timescale threshold on the definition of metastability; it is also different from the role of the tolerance in the dynamic overlap which was discussed in the Introduction.)}

\rev{We find that the overall physics, summarized by the form of the extended phase diagram in the $(\epsilon, T)$ plane, is robust to the choice of the tolerance (or, equivalently, coarse-graining length) $a$. The existence of a first-order transition between a low-overlap phase and a high-overlap one and of a terminal critical point does not depend on the choice of $a$, or equivalently on how exactly similar should two configurations be to belong to the same state. Neither the limit of small $a$ nor that of large $a$ appears to be singular in this respect. This being said, the nonmonotonic dependence of the location of the first-order line (except $T_K$) and of its critical endpoint implies for instance that at a given temperature between $T_K$ and $T_c^{{\rm max}}=\max_{a}\{T_c(a)\}$ there is a range of tolerance $a$ for which one always finds a value of the applied source/coupling at which coexistence between low-overlap and high-overlap phases exists whereas for, the complementary domain of $a$, one is above the critical point and a unique phase is found whatever the applied source. The distribution of overlaps (in the presence of the proper value of the source) is therefore bimodal in the former case and unimodal in the latter. Above $T_c^{{\rm max}}$ there are no more signatures of a  complex landscape, whatever the choice of $a$,  and this temperature is therefore a candidate for a purely static definition, namely one only based on the statistical properties of the configurational space of the liquid, of the ``onset temperature'' below which glassy features starts to set in. Then, the value of $a$ for which this maximum is achieved should represent the typical displacement magnitude of particles in order for the system to fall in another ``metastable'' state, and hence might be rationalized, for instance, by looking at the change in the potential energy in the so-called inherent structure as a function of the mean-squared displacement.}

\rev{The above observation leads us to discuss the notion of configurational entropy as a measure of the number of metastable states, more precisely, of its logarithm divided by the number of particles, in a glass-former. (This definition of the configurational entropy does not exactly correspond to another definition that is associated with the number of minima of the potential-energy hypersurface\cite{stillinger1982hidden,stillinger1995topographic,goldstein1969viscous}.) In a mean-field setting, as already mentioned, the configurational entropy is obtained as the difference in the Franz-Parisi (FP) potential between the metastable and the stable minima, for temperatures between $T_d$ and $T_K$. Above $T_d$, a metastable minimum only occurs in rare instances, no longer in typical ones, and the situation can be described through a large-deviation function\cite{franz2020large}. In any case, there are no ambiguities in counting metastable states from the properties of the minima. As also already discussed, the configurational entropy is then independent of the choice of $a/\sigma$. On the other hand, in a $3$-dimensional glass-former in the thermodynamic limit, there are no $T_d$ and no metastable minimum. However, as seen in Fig.~\ref{Fig_sketch}, there is still a special point, the high-overlap limit of the straight segment, that can serve as a proxy for the latter, and one can tentatively define a configurational entropy as the difference in FP potential between this point and the stable minimum: for instance, see \revminor{Refs.~[\onlinecite{berthier2017configurational},\onlinecite{biroli2016role}]}. What our work shows is that this estimate of the configurational entropy depends on the choice of $a/\sigma$, because the nature or the size of the states that are counted may change with it. In other words, the FP potential is no longer singular in finite dimensions and the choice of the overlap value to compute the difference in FP potential with respect to the stable minimum is now $a$-dependent. One may invoke at this point a physical constraint to fix the value of $a$ and provide the ``most reasonable'' counting by setting $a$ equal to some typical  vibrational length (as obtained, for instance, from the height of the plateau in the mean-squared displacement, at least at temperatures for which the plateau is indeed observed). However, and as practical measurements of the configurational entropy in computer simulations of glass-forming liquid models are currently an important research topic\cite{berthier2017configurational,ozawa2018configurational,ozawa2019does,berthier2017configurational,berthier2014novel,ozawa2017does,berthier2019configurational,berthier2019zero}, the question should certainly be investigated more thoroughly.}
\\

\begin{acknowledgments}
We thank Chiara Cammarota and Francesco Zamponi for providing the trial functions and their code to solve the HNC equations for the bulk liquid.
B. Guiselin acknowledges support by Capital Fund Management - Fondation pour la Recherche. This work was supported by a grant from the Simons Foundation (Grant No. 454933, L.B.).
\end{acknowledgments}

\section*{Data availability}
The data that support the findings of this study are available from the corresponding author upon reasonable request.

\appendix

\section{Analysis of the HNC equations when $a\to 0^+$}
\label{app_analysisHNC}

\begin{figure}[!t]
 \begin{center}
\includegraphics[width=\columnwidth]{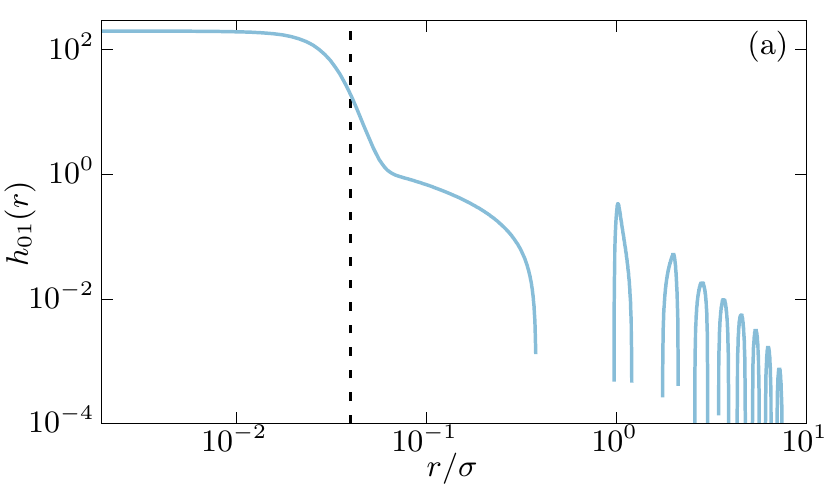} 
  \includegraphics[width=\columnwidth]{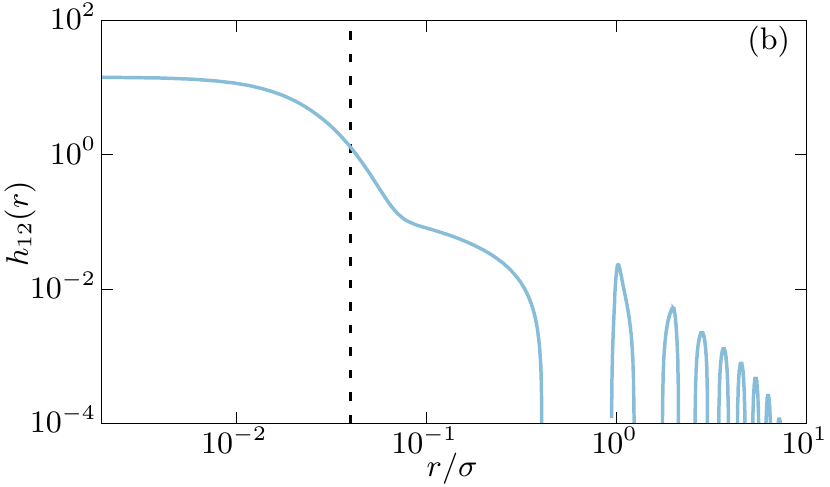} 
\caption{Log-log plot of the HNC total correlation functions $h_{01}$ (a) and $h_{12}$ (b) versus $r/\sigma$ at criticality for $a/\sigma=0.04$ in the case of hard spheres. Notice the decoupling of scales between the range $r\sim a$ where a monotonic decrease is observed and the range $r\sim \sigma$ where oscillations due to the underlying liquid structure occur (the dashed line marks $r=a$).}
\label{Fig_small-a_h}
\end{center}
\end{figure}

When $a/\sigma\to 0^+$ the correlation functions should vary on two very different scales. On the scale $r/\sigma$ one expects a perturbation of the case $a=0$ whereas a singular behavior should appear on the scale $r/a$.  One then considers the following ansatz  at the critical point (for convenience we omit the subscript $c$ on all the quantities evaluated at this critical point):
\begin{equation}
\begin{aligned}
 \label{eq_small-a_ansatz_h}
&h_{01}(r)= \hat f_1(a)\hat h_{01}(r/a)+\tilde f_1(a) \tilde h_{01}(r/\sigma),\\&
h_{12}(r)= \hat f_2(a) \hat h_{12}(r/a)+\tilde f_2(a) \tilde h_{12}(r/\sigma),\\&
h_{11}(r)=\tilde h_{11}(r/\sigma)=\tilde h_{11}^{(0)}(r/\sigma)+\tilde f_3(a)\tilde h_{11}^{(1)}(r/\sigma),\\&
h_{00}(r)=\tilde h_{00}(r/\sigma),
\end{aligned}
\end{equation}
where the hat and tilde functions $h$ have an amplitude and a range of $O(1)$ (see Fig.~\ref{Fig_small-a_h}). Except for $\tilde h_{00}$ which is independent of $a$, they could still have subdominant terms in $a$ as we have shown explicitly for $\tilde h_{11}$. 

In Fourier space the above expressions translate into
\begin{equation}
\begin{aligned}
 \label{eq_small-a_ansatzFourier_h}
&h_{01}(q)= a^{3}\hat f_1(a)\hat h_{01}(qa)+ \tilde f_1(a)\sigma^{3}\tilde h_{01}(q\sigma),\\&
h_{12}(q)= a^{3}\hat f_2(a)\hat h_{12}(qa)+\tilde f_2(a) \sigma^{3} \tilde h_{12}(q\sigma),\\&
h_{11}(q)= \sigma^{3}\tilde h_{11}^{(0)}(q\sigma)+\tilde f_3(a) \sigma^{3} \tilde h_{11}^{(1)}(q\sigma),\\&
h_{00}(q)= \sigma^{3}\tilde h_{00}(q\sigma),
\end{aligned}
\end{equation}
where for simplicity we keep the same notation for the functions in real and Fourier spaces. 

We expect that the prefactors expressing the dependence on $a\to 0^+$ satisfy 
\begin{equation}
\begin{aligned}
&\hat f_1(a),\; \hat f_2(a)\to +\infty,\\&
a^3\hat f_1(a),\; a^3\hat f_2(a)\to 0,\\&
 \tilde f_1(a),\; \tilde f_2(a),\; \tilde f_3(a)\to 0.
\end{aligned}
\end{equation}
Recall also that the function $h_{{\rm con}}$ is the difference between $h_{11}$ and $h_{12}$. The tilde functions varying on the scale $\sigma$ should keep track of the liquid structure and peak in Fourier space around $2\pi/\sigma$. On the other hand, the hat functions are expected to behave roughly as the function $w$ and decay in Fourier space on a scale $q\sim 1/a$: see Fig.~\ref{Fig_small-a_h}. As a result, a complete separation of scales between the hat and tilde functions requires $2\pi/\sigma \ll 1/a$. This is of course verified in the limit $a\to 0^+$ but is more difficult to achieve in the numerical solution of the HNC equations: for instance, when $a/\sigma=0.06$, $2\pi a/\sigma$ is still about  $0.38$, which is smaller but not much smaller than $1$, and corrections to the asymptotic analysis of the functions should then be expected.

By using the separation of the scales $a$ and $\sigma$ the HNC closure in Eqs.~(\ref{eq_HNC_closure}) then leads to direct correlation functions that have a similar structure as their counterparts in Eqs.~(\ref{eq_small-a_ansatz_h}). They are given at the first dominant orders by
\begin{equation}
\begin{aligned}
 \label{eq_small-a_ansatz_c}
&c_{01}(r)= \left\lbrace \hat f_1(a)\hat h_{01}(r/a)-\ln[1+ \hat f_1(a)\hat h_{01}(r/a)]+ \right.\\&
\left.\hat f_3(a)\widehat{\beta\epsilon} w(r/a)\right\rbrace+ \tilde f_1(a)^2 \tilde c_{01}(r/\sigma),\\&
c_{12}(r)= \left\lbrace \hat f_2(a)\hat h_{12}(r/a)-\ln[1+ \hat f_2(a)\hat h_{12}(r/a)]\right\rbrace\\&
+ \tilde f_2(a)^2 \tilde c_{12}(r/\sigma),\\&
c_{11}(r)=\tilde c_{11}^{(0)}(r/\sigma)+ \tilde f_3(a) \tilde c_{11}^{(1)}(r/\sigma),\\&
c_{00}(r)=\tilde c_{00}(r/\sigma),
\end{aligned}
\end{equation}
where $\tilde c_{01}(r/\sigma)=\tilde h_{01}(r/\sigma)^2/2$, $\tilde c_{12}(r/\sigma)=\tilde h_{12}(r/\sigma)^2/2$, $\tilde c_{11}^{(0)}(r/\sigma)=-\beta v(r)+\tilde h_{11}^{(0)}(r/\sigma)-\ln[1+\tilde h_{11}^{(0)}(r/\sigma)]$, $\tilde c_{11}^{(1)}(r/\sigma)=\tilde h_{11}^{(1)}(r/\sigma)\tilde h_{11}^{(0)}(r/\sigma)/[1+\tilde h_{11}^{(0)}(r/\sigma)]$,  $\tilde c_{00}(r/\sigma)=-\beta v(r)+\tilde h_{00}(r/\sigma)-\ln[1+\tilde h_{00}(r/\sigma)]$, and we have assumed that, at criticality, when $a\to 0^+$,
\begin{equation}
\beta\epsilon=\hat f_3(a)\widehat{\beta\epsilon}\; \;\;{\rm with}\;\, \hat f_3(a)\to +\infty. 
\end{equation}

As mentioned above, the functions $\hat h_{01}(r/a)$ and $\hat h_{12}(r/a)$ are expected to behave roughly as $w(r/a)$, {\it i.e.} to decay essentially monotonically on a scale of $O(1)$. As a result, one can rewrite 
\begin{equation}
\begin{aligned}
 \label{eq_small-a_ansatz_h_log}
&\ln[1+ \hat f_1(a)\hat h_{01}(r/a)]\approx \ln [\hat f_1(a)] \hat F_1(r/a) , \\&
\ln[1+ \hat f_2(a)\hat h_{01}(r/a)]\approx \ln [\hat f_2(a)] \hat F_2(r/a),
\end{aligned}
\end{equation}
where the functions $ \hat F_{1,2}$ have an amplitude and a range of $O(1)$ [{\it e.g.} if the function, say, $\hat h_{01}(r/a)$ is approximated by a step function, the function $\hat F_1(r/a)$ verifies $\hat F_1(r/a)\approx \hat h_{01}(r/a)/\hat h_{01}(0)$].

In Fourier space, the expressions in Eqs.~(\ref{eq_small-a_ansatz_c}) become
\begin{equation}
\begin{aligned}
 \label{eq_small-a_ansatzFourier_c}
&c_{01}(q)= \left\lbrace a^{3}\hat f_1(a) \hat h_{01}(qa)-a^3 \ln [\hat f_1(a)] \hat F_1(qa)+\right.\\&
\left. a^{3}\hat f_3(a)\widehat{\beta\epsilon}\, w(qa)\right\rbrace + \tilde f_1(a)^{2} \sigma^{3}\tilde c_{01}(q\sigma),\\&
c_{12}(q)= \left\lbrace a^{3}\hat f_2(a)\hat h_{12}(qa)-a^3 \ln [\hat f_2(a)] \hat F_2(qa)\right\rbrace\\&
+\tilde f_2(a)^{2} \sigma^{3} \tilde c_{12}(q\sigma),\\&
c_{11}(q)= \sigma^{3}\tilde c_{11}^{(0)}(q\sigma)+\tilde f_3(a)\sigma^3\tilde c_{11}^{(1)}(q\sigma),\\&
c_{00}(q)= \sigma^{3}\tilde c_{00}(q\sigma)\,.
\end{aligned}
\end{equation}

We now consider the Ornstein-Zernike equations [Eq.~(\ref{eq_HNC_Fourier})] that can be studied for $q\sim1/a\gg2\pi/\sigma$ and for $q\sim 2\pi/\sigma \ll 1/a$ separately. The decoupling between the two scales in Fourier space is achieved when $a\to 0^+$ if all the $\tilde h_{\alpha\gamma}(q\sigma)$'s go to zero fast enough when $q\sigma \to \infty$: one expects that they indeed do so at least as fast as $1/(q\sigma)^2$ (as in the Ornstein-Zernike approximation).

\begin{figure}[!t]
 \begin{center}
   \includegraphics[width=\columnwidth]{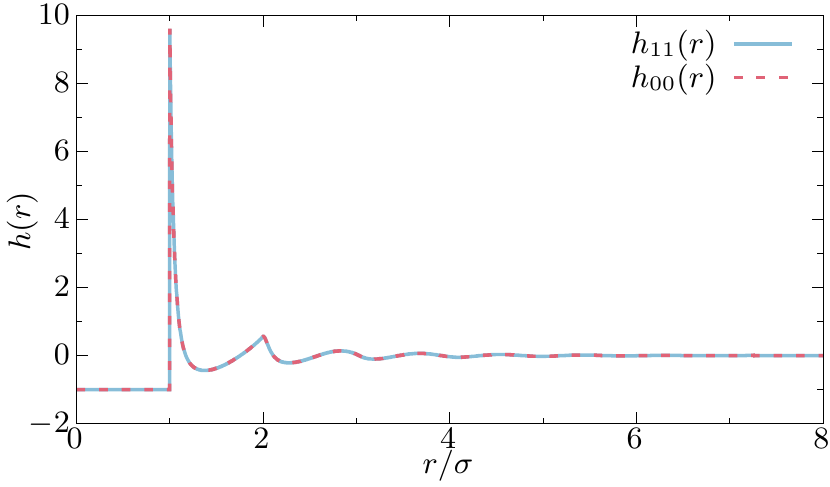}  
\caption{HNC total correlation functions $h_{11}$ and $h_{00}$ versus $r/\sigma$ at criticality for $a/\sigma=0.04$ in the case of hard spheres. The two functions nearly coincide.}
\label{Fig_h00-h11}
\end{center}
\end{figure}

The relation between $h_{01}(q)$ and $c_{01}(q)$ reads
\begin{equation}
\begin{aligned}
 \label{eq_small-a_h01}
&a^{3}\hat f_1(a)\hat h_{01}(qa)+ \tilde f_1(a)\sigma^{3}\tilde h_{01}(q\sigma)=[1+ \sigma^{3}\rho\tilde h_{00}(q\sigma)]\times \\&
[1+\sigma^{3}\rho\tilde h_{11}^{(0)}(q\sigma)+\tilde f_3(a) \sigma^{3}\rho \tilde h_{11}^{(1)}(q\sigma)-a^{3}\hat f_2(a) \rho \hat h_{12}(qa) -\\&
\tilde f_2(a)\sigma^{3} \rho \tilde h_{12}(q\sigma)]\lbrace a^{3}\hat f_1(a) \hat h_{01}(qa)-a^3 \ln [\hat f_1(a)] \hat F_1(qa)+\\&
a^{3}\hat f_3(a)\widehat{\beta\epsilon}\, w(qa) +{\rm O}(\tilde f_1(a)^{2})\rbrace.
\end{aligned}
\end{equation}
When $q\sim 1/a$, one can neglect the contributions of the tilde functions and the above equation implies that
\begin{equation}
\begin{aligned}
\label{eq_small-a_hath01}
&a^{3}\hat f_1(a)\hat f_2(a)\rho \hat h_{12}(qa)\hat h_{01}(qa) + \ln [\hat f_1(a)] \hat F_1(qa)=\\&\hat f_3(a)\widehat{\beta\epsilon}\, w(qa).
\end{aligned}
\end{equation}
On the other hand, when $q\sim 2\pi/\sigma$, one has
\begin{equation}
\begin{aligned}
 \label{eq_small-a_h01_1}
&a^{3} \hat f_1(a)\hat h_{01}(q=0)\lbrace 1-[1+\sigma^{3}\rho \tilde h_{00}(q\sigma)][1+\sigma^{3}\rho\tilde h_{11}^{(0)}(q\sigma)]\rbrace\\&
+\tilde f_1(a) \sigma^{3}\tilde h_{01}(q\sigma)={\rm o}(a^{3} \hat f_1(a),\tilde f_1(a)),
\end{aligned}
\end{equation}
where the right-hand side only contains terms that are subdominant compared to $a^{3} \hat f_1(a)$ and/or $\tilde f_1(a)$. The above equation therefore implies that
\begin{equation}
\begin{aligned}
\label{eq_tildeh01}
\tilde f_1(a)=a^{3} \hat f_1(a)
\end{aligned}
\end{equation}
and that
\begin{equation}
\begin{aligned}
\label{eq_tildeh01bis}
&\sigma^3 \tilde h_{01}(q\sigma)=\\&\hat h_{01}(q=0)\lbrace[1+\sigma^{3}\rho \tilde h_{00}(q\sigma)][1+\sigma^{3}\rho\tilde h_{11}^{(0)}(q\sigma)]-1\rbrace,
\end{aligned}
\end{equation}
with an unimportant choice of normalization of the functions.

We now proceed in a similar way for the Ornstein-Zernike equation that relates $h_{12}(q)$ and $c_{12}(q)$. It reads
\begin{equation}
\begin{aligned}
 \label{eq_small-a_h12}
&a^{3}\hat f_2(a)\hat h_{12}(qa)+\tilde f_2(a) \sigma^{3} \tilde h_{12}(q\sigma)=[1+\sigma^{3}\rho\tilde h_{11}^{(0)}(q\sigma)+ \\&
\tilde f_3(a) \sigma^{3}\rho \tilde h_{11}^{(1)}(q\sigma)-a^{3}\hat f_2(a)\rho \hat h_{12}(qa) -\tilde f_2(a) \sigma^{3} \rho \tilde h_{12}(q\sigma)]^2 
\\&\times \lbrace a^{3}\hat f_2(a) \hat h_{12}(qa)-a^3 \ln [\hat f_2(a)] \hat F_2(qa)+{\rm O}(\tilde f_2(a)^2)+\\&
\rho [1+ \sigma^{3}\rho\tilde h_{00}(q\sigma)] [a^{3}\hat f_1(a)\hat h_{01}(q a)+\cdots]^2\rbrace,
\end{aligned}
\end{equation}
where $\cdots$ denotes terms that, following Eqs.~(\ref{eq_small-a_hath01}), (\ref{eq_tildeh01}) and (\ref{eq_tildeh01bis}), are subdominant compared to $a^{3}\hat f_1(a)$.

\begin{figure}[!t]
 \begin{center}
 \includegraphics[width=\columnwidth]{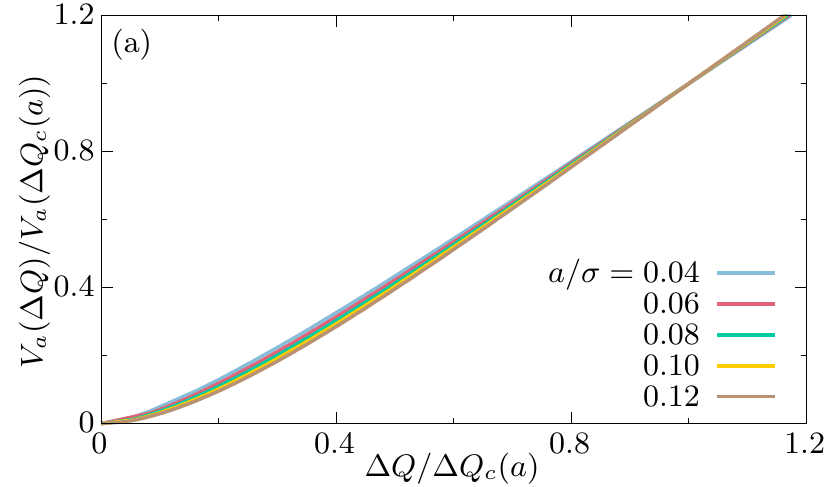} 
 \includegraphics[width=\columnwidth]{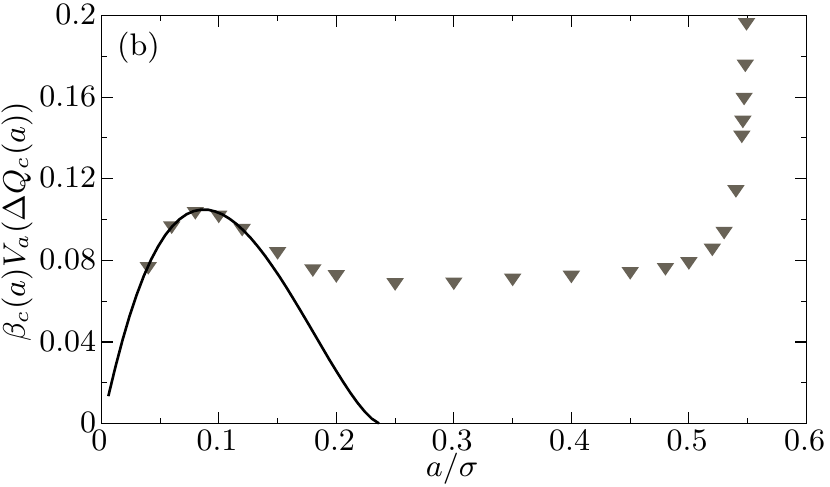} 
\caption{Rescaled FP potential in the limit $a\to 0^+$ for hard spheres in the HNC approximation: (a) $V_a(\Delta Q)/V_a(\Delta Q_c(a))$ versus $\Delta Q/\Delta Q_c(a)$ for several values of $a\leq 0.12$; (b) $V_a(\Delta Q_c(a))$ versus $a$: when $a$ decreases the value first passes through a maximum but then steadily decreases in a manner compatible with the prediction $(a\vert \ln a\vert)^{3/2}$ (full line).}
\label{Fig_app_rescaledFP}
\end{center}
\end{figure}

When $q\sim 1/a$, one can neglect the contributions of the tilde functions again and one finds
\begin{equation}
\begin{aligned}
\label{eq_small-a_hath12}
a^{3}\hat f_1(a)^2\rho \hat h_{01}(qa)^2 &= 2a^{3}\hat f_2(a)^2\rho \hat h_{12}(qa)^2\\&
+\ln [\hat f_2(a)] \hat F_2(qa).
\end{aligned}
\end{equation}
When $q\sim 2\pi/\sigma$, after using some of the already obtained relations, one obtains
\begin{equation}
\begin{aligned}
 \label{eq_small-a_h12_1}
&a^{3}\hat f_2(a) \hat h_{12}(q=0)\lbrace 1-[1+\sigma^{3}\rho\tilde h_{11}^{(0)}(q\sigma)]^2\rbrace +\\&
\tilde f_2(a) \sigma^{3}\tilde h_{12}(q\sigma)= {\rm o}(a^{3} \hat f_2(a),\tilde f_2(a)),
\end{aligned}
\end{equation}
where the right-hand side is subdominant compared to $a^{3} \hat f_2(a)$ and $\tilde f_2(a)$. This implies that 
\begin{equation}
\begin{aligned}
\label{eq_tildeh12}
\tilde f_2(a)=a^{3} \hat f_2(a),
\end{aligned}
\end{equation}
and that
\begin{equation}
\begin{aligned}
\label{eq_tildeh12bis}
&\sigma^3 \tilde h_{12}(q\sigma)=\\&\hat h_{12}(q=0)\lbrace[1+\sigma^{3}\rho\tilde h_{11}^{(0)}(q\sigma)]^2-1\rbrace,
\end{aligned}
\end{equation}
with an unimportant choice of normalization of the functions.

Although Eqs.~(\ref{eq_small-a_hath01}) and (\ref{eq_small-a_hath12}) could have several possible solutions, a nontrivial solution is obtained by assuming that in each of these equations all terms are of the same order. This gives
\begin{equation}
\begin{aligned}
&a^{3}\hat f_1(a)\hat f_2(a)\sim \hat f_3(a) \sim \ln[\hat f_1(a)],\\&
a^{3}\hat f_1(a)^2 \sim a^3 \hat f_2(a)^2\sim \ln[\hat f_2(a)],
\end{aligned}
\end{equation}
whose solution is then, at leading order when $a\to 0^+$,
\begin{equation}
\begin{aligned}
&\hat f_1(a)\sim \hat f_2(a)\sim a^{-3/2} \sqrt{\vert\ln a\vert},\\&
\hat f_3(a)\sim \vert \ln a\vert.
\end{aligned}
\end{equation}

Finally, we consider the Ornstein-Zernike equation relating $h_{{\rm con}}(q)$ and $c_{{\rm con}}(q)$:
\begin{equation}
\begin{aligned}
 \label{eq_small-a_hcon}
&1+\sigma^{3}\rho\tilde h_{11}^{(0)}(q\sigma)+\tilde f_3(a)\sigma^{3}\rho \tilde h_{11}^{(1)}(q\sigma)-\tilde f_2(a)\rho [\hat h_{12}(qa) + \\&
\sigma^{3} \tilde h_{12}(q\sigma)]=\left\lbrace 1-\sigma^{3} \rho\tilde c_{11}^{(0)}(q\sigma)-\tilde f_3(a)\sigma^{3}\rho\tilde c_{11}^{(1)}(q\sigma) + \right.\\&
\left.\tilde f_2(a)\rho\hat h_{12}(qa)-a^3\ln[\hat f_2(a)]\rho\hat F_2(qa) + {\rm O}(\tilde f_2(a)^2)\right\rbrace^{-1},
\end{aligned}
\end{equation}
where we have used Eq.~(\ref{eq_tildeh12}). At leading order this immediately leads to
\begin{equation}
1+\sigma^{3}\rho\tilde h_{11}^{(0)}(q\sigma)=\frac 1{1-\sigma^{3}\rho\tilde c_{11}^{(0)}(q\sigma)},
\end{equation}
and since the HNC closures for $\tilde c_{11}^{(0)}$ and $\tilde c_{00}$ have the same form, to
\begin{equation}
\label{eq_tildeh11}
\tilde h_{11}^{(0)}(q\sigma)=\tilde h_{00}(q\sigma),
\end{equation}
which is well verified by our numerical solution of the HNC equations (see Fig.~\ref{Fig_h00-h11}). 

In addition, by using Eq.~(\ref{eq_tildeh11}) as well as Eq.~(\ref{eq_tildeh12bis}), one finds at the next-to-leading orders and when $q\sim 2\pi/\sigma$ that
\begin{equation}
\begin{aligned}
 \label{eq_small-a_hcon_1}
&\tilde f_3(a)\sigma^3\lbrace \tilde h_{11}^{(1)}(q\sigma)-[1+\sigma^3\rho\tilde h_{00}(q\sigma)]^2\tilde c_{11}^{(1)}(q\sigma)\rbrace=\\&
a^3\ln[\hat f_2(a)]\hat F_2(q=0)[1+\sigma^3\rho\tilde h_{00}(q\sigma)]^2\,.
\end{aligned}
\end{equation}
Assuming that the terms on both sides of the equation are of the same order, Eq.~(\ref{eq_small-a_hcon_1}) leads to
\begin{equation}
\label{eq_tildehf3}
\tilde f_3(a)=a^{3} \ln[\hat f_2(a)]\sim a^3 \vert\ln a\vert,
\end{equation}
at the leading order when $a\rightarrow 0^+$, and
\begin{equation}
\sigma^3\tilde h_{11}^{(1)}(q\sigma)=[1+\sigma^3\rho\tilde h_{00}(q\sigma)]^2[\sigma^3\tilde c_{11}^{(1)}(q\sigma) +
\hat F_2(q=0)]\,.
\end{equation}
The above derivation provides the expressions given in the main text.

To conclude this appendix, we consider the FP potential and assume it follows a scaling form when $a\to 0^+$,
\begin{equation}
\label{eq_scaling-FPpot}
\frac{V_a(\Delta Q)}{V_a(\Delta Q_c(a))} \to \phi(\frac{\Delta Q}{\Delta Q_c(a)}),
\end{equation}
with $\phi(x)$ a scaling function. (As usual we consider the FP potential to be zero at the absolute minimum corresponding to decoupled replicas.) This scaling behavior is indeed supported by the data in Fig.~\ref{Fig_app_rescaledFP}(a).
By definition of the critical point, $\phi''(1)=\phi'''(1)=0$ and $(\beta_c\epsilon_c)(a)=\phi'(1)V_a(\Delta Q_c(a))/\Delta Q_c(a)$ [see Eqs.~(\ref{eq_critical1}) and (\ref{eq_critical2})]. From the behavior of $(\beta_c\epsilon_c)(a)$ and $\Delta Q_c(a)$ (see the main text), one then predicts that $V_a(\Delta Q_c(a))\sim (\beta_c\epsilon_c)(a)\Delta Q_c(a)$ goes to zero as $a\to 0^+$ as $(a\vert \ln a\vert)^{3/2}$. This is compatible with the data in Fig.~\ref{Fig_app_rescaledFP}(b).

\bibliography{biblio_HNC.bib}

\end{document}